\documentclass[12pt,letterpaper]{article}
\usepackage{amsmath,amssymb}
\usepackage{graphicx,color}
\usepackage[linkcolor=blue,colorlinks=true]{hyperref}
\usepackage{cite,epsf}
\usepackage[bf]{caption}
\usepackage{subfig}
\usepackage[nottoc,notlot,notlof]{tocbibind}

\usepackage{tikz}
\usetikzlibrary{positioning}
\usetikzlibrary{intersections} 
\newlength{\PicScale}
\newcommand{\rr}{\mathbb{R}}
\newcommand{\zz}{\mathbb{Z}}
\newcommand{\sz}{\mathbb{S}}

\newcommand{\be}{\begin{equation}}
\newcommand{\ee}{\end{equation}}
\newcommand{\ba}{\begin{aligned}}
\newcommand{\ea}{\end{aligned}}
\newcommand{\ben}{\begin{displaymath}}
\newcommand{\een}{\end{displaymath}}
\newcommand{\bea}{\begin{eqnarray}}
\newcommand{\eea}{\end{eqnarray}}
\newcommand{\bean}{\begin{eqnarray*}}
\newcommand{\eean}{\end{eqnarray*}}
\newcommand{\f}{\frac}
\newcommand{\p}{\partial}

\renewcommand{\theequation}{\thesection.\arabic{equation}}

\def\th {\theta}
\def\a {\alpha}

\def\b {\beta}

\def\g {\gamma}
\def\G {\Gamma}
\def\d {\delta}
\def\e {\epsilon}
\def\s {\sigma}
\def\e {\epsilon}
\def\z {\zeta}

\def\m{\mu}
\def\n{\nu}

\def\o{\omega}

\definecolor{green}{rgb}{0,0.5,0}

\def\p{\partial}

\usepackage{epsfig}

\usepackage{xargs}
\usepackage[colorinlistoftodos,prependcaption,textsize=tiny]{todonotes}
\newcommandx{\Stefan}[1]{\todo[backgroundcolor=red!25,bordercolor=red,noline]{S:#1}}
\newcommandx{\George}[1]{\todo[backgroundcolor=blue!25,bordercolor=blue,noline]{G:#1}}

\numberwithin{equation}{section}

\begin{document}

\begin{titlepage}       \vspace{10pt} \hfill 

\vspace{20mm}

\begin{center}

{\large \bf On the S-matrix of Liouville theory}

\vspace{30pt}

George Jorjadze$^{a,\,b,\,c}~$ and Stefan Theisen$^{c}$ 
\\[6mm]

{\small
{\it ${}^a$Free University of Tbilisi,\\
		Agmashenebeli Alley 240, 0159, Tbilisi, Georgia}\\[2mm]
{\it ${}^b$Razmadze Mathematical Institute of TSU,\\
Tamarashvili 6, 0177, Tbilisi, Georgia}\\[2mm]
{\it ${}^c$Max-Planck-Institut f\"ur Gravitationsphysik, Albert-Einstein-Institut,\\ 
14476, Golm, Germany}
}

\vspace{20pt}

\end{center}

\centerline{{\bf{Abstract}}}
\vspace*{5mm}
\noindent

The $S$-matrix for each chiral sector of Liouville theory on a cylinder is computed from the loop expansion of correlation functions 
of a one-dimensional field theory on a circle with a non-local kinetic energy and an exponential potential. This action is the 
Legendre transform of the generating function of semiclassical scattering amplitudes. It is derived from the relation between
asymptotic $in$- and $out$-fields. Its relevance for the quantum scattering process is demonstrated by comparing 
explicit loop diagrams computed from this action with other methods of computing the $S$-matrix, which are also 
developed.

\vspace{15pt}

\end{titlepage}

\tableofcontents{}
\vspace{1cm}
\bigskip\hrule

\section{Introduction and summary}

Liouville theory is one of the most interesting and most studied interacting two-dim\-ension\-al 
conformal field theories. It plays a central role in
non-critical bosonic string theory where it arises when one integrates the Weyl anomaly of the Polyakov action in a non-trivial 
metric background in conformal gauge \cite{Polyakov:1981rd}.  There are several reviews which cover the subject very well  and 
we refer to them for further motivations and thorough discussions \cite{Teschner:2001rv,Nakayama:2004vk,Seiberg:1990eb}. 
Various generalisations have been considered, 
in particular supersymmetric ones, but also to higher dimensions \cite{Levy:2018bdc}.  
Here we will only discuss the two-dimensional bosonic theory.   
We concentrate on a particular aspect of Liouville theory, namely the $S$-matrix of scattering states, 
a subject which was initiated in \cite{Zamolodchikov:1995aa}.  An expression for the $S$-matrix in closed form 
is not known and the purpose of this note is to report an observation which allows for a reformulation of the problem, 
which might shed a new light on the issue. 

The dynamics of Liouville theory is derived from the Liouville action. The classical equations of motion can be solved 
and the general solution is specified by two arbitrary functions. They are defined, respectively, on the two branches of 
the light cone or, in the Euclidean formulation of the theory, by an holomorphic and an 
anti-holomorphic function. These functions have to respect some regularity conditions and, 
as we will consider the theory on a cylinder, a periodicity requirement as well.   
The only regular solutions on the cylinder are scattering solutions which interpolate between free asymptotic $in$- and 
free asymptotic $out$-states. The transformation from $in$- to $out$-states is effected by the $S$-matrix. 

Our aim in this paper is to compute the quantum $S$-matrix of Liouville theory on the cylinder with 
Minkowski signature. We proceed in two steps.
In the first step we compute the semiclassical $S$-matrix or, more precisely, the generating functional ${\cal F}$ for semiclassical 
$S$-matrix elements in the Fock-space generated by the oscillator modes. 
It is the generating function of the canonical transformation between the $in$- and $out$-fields which are free fields and 
therefore split into two chiral sectors.  We show that this transformation is canonical for each chiral sector separately, 
with generating functionals $F$ and $\bar F$. The two sectors are only coupled through the zero-modes.  
In terms of the oscillator modes of the $in$- and $out$-fields, in each sector it is a function of the 
positive modes of the former and of the negative modes of the latter and, of course, of the zero mode, i.e. the momentum.
As we will explain in detail, the positive and negative modes in each chiral sector form conjugate pairs 
which are related by a Legendre transformation with generating function $F$,  
and, as usually, the momentum is conjugate to the position.    

A slight but crucial modification of $F$, which affects only the two-point functions, leads to a re-organization of the 
conjugate pairs in such a way that we can compute its Legendre transformation. This is essentially just 
a rewriting of the known relation between the $in$- and $out$-fields. 
The Legendre transform, which we call  $S$, has the form of a one-dimensional classical action with an unusual 
non-local kinetic energy term and an exponential potential and where 
the momentum now plays the role of a coupling constant. 
We claim that this action, when quantized, contains all the information 
about the quantum scattering in one chiral sector.
A priori this is not clear as the relevance of this action for the quantum theory is not obvious. 
We support our claim by first computing 
loop diagrams, where the only regularization required is `normal ordering' of the potential. We compare these loop results 
with those obtained from another way to compute scattering amplitudes. This method, which works for the 
semiclassical and the quantum case, uses the fact that the improved energy-momentum tensors
for the $in$- and $out$-fields in each chiral sector coincide. Equating them and expressing the resulting equation in 
terms of the independent canonical variables, leads to a differential equation for the generating functional which can be solved 
in a power series in the modes. We use this second method, which for the quantum amplitudes is more efficient than 
the loop computations and produces results which are valid to all powers in $\hbar$, to check the Feynman diagram 
computations.  We also find agreement with the results obtained by Zamolodchikov \& Zamolodchikov in \cite{Zamolodchikov:1995aa}. 

The organization of the paper is as follows. 
In Section \ref{dynamics} we review known facts about classical Liouville theory on the cylinder, which will be relevant for the 
subsequent discussion. Next, in Section \ref{Canonical Structure}, we discuss the 
canonical structure and introduce generating functions for canonical transformations 
between the $in$- and $out$-fields. The most relevant is the transformation between $in$- and $out$-oscillator modes, 
which is canonical in each chiral sector separately. Its generating function $F$ completely specifies the semiclassical
$S$-matrix in one chiral sector.  Further details about the canonical structure are given in two appendices.  
$F$ will be computed in Section \ref{semi}, where the relation between the classical 
$in$- and $out$-fields is used to find $S$, the Legendre transform of $F$. 
We use it to compute semiclassical scattering amplitudes via tree-level 
Feynman diagrams derived from $S$ and compare them with a direct computation, which exploits the equality between the 
improved energy-momentum tensors of the asymptotic $in$- and $out$-states. In Section \ref{quantum} we 
compute quantum corrections to the scattering amplitudes. We first do this by evaluating (mostly) one-loop Feynman diagrams  
and then compare them to an explicit computation which again uses 
the equality of the $in$ and $out$  energy-momentum tensors, with the quantum corrections to the improvement term taken 
into account. We end with short conclusions.

\section{Liouville dynamics and asymptotic fields}\label{dynamics}

We study Liouville theory on a cylinder with spacetime coordinates 
$\s\in \sz^1$ and $\tau\in \rr^1$. The classical theory is described by the Liouville equation 
\be\label{L-eq}
\p^2_{\tau}\Phi-\p^2_{\s}\Phi+4\,\m^2\,{e}^{2\Phi}=0~.
\ee
$\Phi$ is called the Liouville field and $\m^2>0$ is a `cosmological' constant. 

We also introduce the Liouville field exponential $V={e}^{-\Phi}$ and
the stress tensor components, which in chiral coordinates $(x=\tau+\s,~ \bar x=\tau-\s)$ are defined as
\be\label{stress tensor} 
T=(\p_x\Phi)^2-\p^2_{x}\Phi~, \qquad \bar T=(\p_{\bar x}\Phi)^2-\p^2_{\bar x}\Phi~.
\ee
The chirality conditions $\p_{\bar x}T=0,$ $\p_x\bar T=0$ follow from the Liouville equation \eqref{L-eq}. 
By \eqref{stress tensor} $V$ satisfies the pair of Hill equations
\be\label{Hill eq}
\p^2_{x}V(x,\bar x) =T(x)\, V(x,\bar x)~, \qquad  \p^2_{\bar x}V(x,\bar x)=\bar T(\bar x)\,V(x,\bar x)~,
\ee
while \eqref{L-eq}, in addition, leads to 
\be\label{L-eq1}
V\,\p_{x}\p_{\bar x}V-\p_x V\,\p_{\bar x}V=\m^2.
\ee
Setting $T=\bar T=\f{1}{4}\,p^2>0$,  one finds a simple solution of \eqref{Hill eq} and \eqref{L-eq1}
\be\label{V_0}
V=e^{-\Phi}=\f{\m}{p}\left(e^{-p\tau}+e^{p\tau}\right)\,, 
\ee
with $p=\sqrt{p^2}$ and  $\m=\sqrt{\m^2}$.
The corresponding $\s$-independent Liouville field $\Phi$ describes particle dynamics in an exponential potential 
and $p$ is interpreted as the momentum of the {\it{in}}-coming particle, while the momentum of the {\it out}-going particle is $-p$.

The space of solutions of \eqref{L-eq1} is invariant under conformal transformations
\be\label{conformal tr}
V(x,\bar x)\mapsto \tilde{V}(x,\bar x)=\left(\zeta'(x)\bar{\zeta}'(\bar x)\right)^{-\f{1}{2}}\,V\left(\zeta(x),\bar{\zeta}(\bar x)\right),
\ee
where the chiral functions are
monotonic  ($\z'>0, \bar\z'>0$) and allow the mode expansion
\be\label{zeta}
\zeta(x)=\zeta_{0}+x+i\sum_{n\neq 0}\f{\zeta_n}{n}\,e^{-inx}~, \qquad
\bar\zeta(\bar x)=\bar\zeta_{0}+\bar x+i\sum_{n\neq 0}\f{\bar\zeta_n}{n}\,e^{-in\bar x}~.
\ee

The  conformal transformations of \eqref{V_0} generate all regular Liouville fields on the cylinder \cite{Balog:1997zz}. 
The general solution can then be written as 
\be\label{V-general}
e^{-\Phi}=e^{-\Phi_{\text{in}}}+e^{-\Phi_{\text{out}}}~,
\ee
where the fields $\Phi_{\text{in}}$ and $\Phi_{\text{out}}$ are parametrised  as follows: 
\be\ba\label{Phi_in_out}
&\Phi_{\text{in}}\,=\,\f{1}{2}\,\log\left[\xi'(x)\,\bar\xi'(\bar x)\right]+\f{\mu}{2}\left[\xi(x)+\bar\xi(\bar x)\right],\\[1mm]
&\Phi_{\text{out}}=\f{1}{2}\,\log\left[\xi'(x)\,\bar\xi'(\bar x)\right]-\f{\mu}{2}\left[\xi(x)+\bar\xi(\bar x)\right],
\ea\ee
with
\be
\xi(x)=\frac{p}{\mu}\,\zeta(x)\,,\qquad \bar\xi(\bar x)=\f p \mu\,\bar\zeta(\bar x)\,.
\ee
The $in$ and $out$ notation will now be justified.

Due to \eqref{zeta}, the non-periodic $\tau$-dependent parts of $\Phi_{\text{in}}$ and $\Phi_{\text{out}}$  are
$p\,\tau$ and $-p\,\tau$, respectively. Since $p>0$, one can neglect $\exp(-\Phi_{\text{out}})$ 
in \eqref{V-general} at $\tau \rightarrow -\infty$  and $\exp(-\Phi_{\text{in}})$  at $\tau \rightarrow +\infty$. 
Hence, the fields $\Phi_{\text{in}}(\tau,\s)$ and $\Phi_{\text{out}}(\tau,\s)$  are interpreted as the asymptotic fields of 
Liouville theory. They are 2d massless free scalar fields, since they are given as the sum of the chiral and antichiral parts
\be\label{chiral decomposition}
~~~~~~~~\Phi_{\text{in}}=\phi_{\text{in}}(x)+\bar\phi_{\text{in}}(\bar x), \qquad \Phi_{\text{out}}
=\phi_{\text{out}}(x)+\bar\phi_{\text{out}}(\bar x), 
\ee
with
\be\ba\label{chiral parts}
&\phi_{\text{in}}(x)\,\,=\f{1}{2}\,\log\left[\xi'(x)\right]+\f{\mu}{2}\,\xi(x),\\[2mm]
&\phi_{\text{out}}(x)=\f{1}{2}\,\log\left[\xi'(x)\right]-\f{\mu}{2}\,\xi(x).
\ea\ee
The antichiral part is similar and it suffices to analyze the chiral part only.

The integration of \eqref{chiral parts} defines $\xi(x)$ through the asymptotic fields
\be\ba\label{Screening charge}
&e^{\,\mu\,\xi(x)}=\m\,A(x), \qquad  &e^{-\mu\,\xi(x)}=\m\,B(x),
\ea\ee
with
\be\ba 
&A(x)=\int_{-\infty}^x \text{d}y\,e^{2\,\phi_{\text{in}}(y)}, \qquad
&B(x)=\int_x^{\infty}\text{d}y\,e^{2\,\phi_{\text{out}}(y)}.
\ea\ee
The integrated free-field exponents $A(x)$ and $B(x)$ are called screening charges. They are conformal scalars similarly 
to the chiral field $\z(x)$. By \eqref{Screening charge} one has the functional relation between the 
asymptotic fields, $\m^2\,A(x)\,B(x)=1$.

Note that the screening charges $A(x)$ and $\bar A(\bar x)$ parametrise the general solution of the Liouville equation
just in the form as was obtained by Liouville \cite{Liouville} (see \eqref{A, bar A parameterisation}).

From \eqref{chiral parts} we also find the chiral $out$-field through the chiral $in$-field and vice versa
\be\label{out=in} 
\phi_{\text{out}}(x)=\phi_{\text{in}}(x)-\log\left[\m\,A(x)\right]~, \qquad \phi_{\text{in}}(x)=\phi_{\text{out}}(x)-\log\left[\m\,B(x)\right]~.
\ee 
These fields have the monodromy $\phi_{\text{in}}(x+2\pi)=\phi_{\text{in}}(x)+\pi\,p$, $\,\,\phi_{\text{out}}(x+2\pi)
=\phi_{\text{out}}(x)-\pi\,p$, and the standard free-field mode expansion
\be\label{FF modes}
\phi_{\text{in}}(x)=q+\f{p\,x}{2}+i\sum_{n\neq 0}\f{a_n}{n}\,e^{-inx}~, \qquad
\phi_{\text{out}}(x)=\tilde q-\f{p\,x}{2}+i\sum_{n\neq 0}\f{b_n}{n}\,e^{-inx}~.
\ee
Thus, the momentum zero modes of the asymptotic fields differ only by their sign,  
while the transformation of the other modes is highly non-trivial, according to \eqref{out=in}.

The stress tensor \eqref{stress tensor} obtained from \eqref{V-general}-\eqref{Phi_in_out}
is given by the Schwarzian derivative
\be\label{stress tensor=}
T(x)=\f{1}{4}\,p^2\,\z'^{\,2}(x)+\f{1}{4}\left(\f{\z''(x)}{\z'(x)}\right)^2-\f 1 2\,\left(\f{\z''(x)}{\z'(x)}\right)',
\ee
and with \eqref{chiral parts} one finds its `improved' free-field form in terms of the asymptotic fields
\be\label{FF form of T}
T(x)=\phi'^{\,2}_{\text{in}}(x)-\phi''_{\text{in}}(x)=\phi'^{\,2}_{\text{out}}(x)-\phi''_{\text{out}}(x)~.
\ee
Hence, the `improved' free-field stress tensors of the $in$ and $out$-fields coincide.  
This will be crucial when we discuss the $S$-matrix.

\section{Canonical structure}\label{Canonical Structure}

A more transparent free-field interpretation of $\Phi_{\text{in}}$ and $\Phi_{\text{out}}$ is obtained in the 
Hamiltonian description defined  by the first-order action
\be\label{H-action}
S[\Phi,\Pi]=\int \text{d}\tau\int_0^{2\pi} \f{\text{d}\sigma}{2\pi}\left[ \Pi\,\dot\Phi-
\left(\frac{1}{2}\,\Pi^2+\frac{1}{2}\left(\partial_\sigma\Phi\right)^2
+2\,\m^2\,e^{2\Phi}\right)
\right]\,,
\end{equation}
where $\Phi(\s)$ and $\Pi(\s)$ are canonically conjugated variables.

The canonical 2-form of Liouville theory induces a symplectic structure on the space of parametrising chiral fields. 
In Appendix \ref{AppA} we consider the parameterisation of the general solution by the $in$-field and obtain  $\Omega=\o+\bar\o$, where 
$\Omega$ is the canonical 2-form, $\o$ is the standard chiral free-field symplectic form for the $in$-field
\be\label{omega-in}
\o=\int_0^{2\pi}\f{\text{d}x}{2\pi}\left[\d\phi'_{\text{in}}(x)\wedge\d\phi_{\text{in}}(x)\right]+
\f 1 2\,\d p\wedge\d\phi_{\text{in}}(0)~,
\ee
and $\bar\o$ is its antichiral counterpart. The mode expansion \eqref{FF modes} then leads to
\be\label{Can 2-form}
\Omega=\int_0^{2\pi}\f{\text{d}\s}{2\pi}\, \d\Pi(\tau, \s)\wedge\d\Phi(\tau,\s)=
\d p_{\text{in}}\wedge \d q_{\text{in}}+i\sum_{n\neq 0}\left(\f{\d a_{-n}\wedge\d a_n}{n}+\f{\d\bar a_{-n}\wedge\d \bar a_n}{n}\right),
\ee
with  $q_{\text{in}}=q+\bar q\,$ and $p_{\text{in}}=p$. 

The right hand side of \eqref{Can 2-form} corresponds to the canonical 2-form of the complete $in$-field
\be\label{in-field}
\Phi_{\text{in}}(\tau,\s)=q_{\text{in}}+p_{\text{in}}\,\tau+i\sum_{n\neq 0}\left(\f{a_n}{n}\,e^{-inx} 
+\f{\bar a_n}{n}\,e^{-in\bar x}\right),
\ee
and its inversion provides the Poisson brackets
\be\label{Can PB}
\{p_{\text{in}},q_{\text{in}}\}=1~, 
\quad \{a_m,a_n\}=\f{i}{2}\,m\,\d_{m,-n}~,\quad \{\bar a_m,\bar a_n\}=\f{i}{2}\,m\,\d_{m,-n}~.
\ee
Thus, the transformation from the $in$-field to the Liouville field is canonical
\be\label{Omega=Omega-in}
\int_0^{2\pi}\f{{\text{d}}\s}{2\pi}\, \d\Pi(\tau, \s)\wedge\d\Phi(\tau,\s)
=\int_0^{2\pi}\f{{\text{d}}\s}{2\pi}\, \d\Pi_{\text{in}}(\tau, \s)\wedge\d\Phi_{\text{in}}(\tau,\s)~.
\ee
The same relation holds for the $out$-field and, therefore, the map from the $in$-field to the $out$-field is canonical as well
\be\label{Omega-in=Omega-out}
\int_0^{2\pi}\f{{\text{d}}\s}{2\pi}\, \d\Pi_{\text{in}}(\tau, \s)\wedge\d\Phi_{\text{in}}(\tau,\s)
=\int_0^{2\pi}\f{\mbox{d}\s}{2\pi}\, \d\Pi_{\text{out}}(\tau, \s)\wedge\d\Phi_{\text{out}}(\tau,\s)~.
\ee

Applying the map \eqref{out=in} to the chiral symplectic form \eqref{omega-in}, we find
\be\label{omega-out}
\o=\int_0^{2\pi}\f{\mbox{d}x}{2\pi}\left[\d\phi'_{\text{out}}(x)\wedge\d\phi_{\text{out}}(x)\right]-
\f 1 2\,\d p\wedge\d\phi_{\text{out}}(0)~.
\ee
Hence, the map is canonical in the chiral sectors separately. 
We will now calculate the corresponding generating functions.

First we consider the generating function $G$ for the canonical transformation in the chiral sector. 
According to \eqref{omega-in} and \eqref{omega-out},  it is  defined by the differential form
\be\label{th_out-th_in}
\d G=\int_0^{2\pi}\f{\mbox{d}x}{2\pi}\,\left[\phi'_{\text{out}}(x)\d\phi_{\text{out}}(x)-\phi'_{\text{in}}(x)\,\d\phi_{\text{in}}(x)\right]+
\f 1 2\,\left[\phi_{\text{out}}(0)+\phi_{\text{in}}(0)\right]\d p~.
\ee
Using the parameterisation \eqref{chiral parts} and partial integration,  we  find $G$ as a function of the chiral part of the phase 
space
\be\label{G=}
G=-p+\f{\m}{2}\int_0^{2\pi}\f{\mbox{d}x}{2\pi}\,\,\xi'\log \xi'\,.
\ee

By \eqref{chiral parts} and \eqref{FF modes}, this function can be written in terms of the Fourier modes as follows
\be\label{G=1}
G=\tfrac{1}{2}p(\tilde q+q)-p+i\sum_{n\neq 0}\f{1}{n}a_{-n}b_{n}~.
\ee

Now we calculate the generating function of the canonical transformation for the complete asymptotic fields. 
Due to \eqref{Omega-in=Omega-out}, the difference between the canonical presymplectic forms of the $out$ and $in$ 
fields is an exact 1-form
\be\label{Phi-Phi}
\d {\cal G}[\Phi_{\text{out}},\Phi_{\text{in}}]=\int_0^{2\pi}\f{\mbox{d}\s}{2\pi}\,
\left[ \Pi_{\text{out}}(\tau, \s)\,\d\Phi_{\text{out}}(\tau,\s)-\Pi_{\text{in}}(\tau, \s)\,\d\Phi_{\text{in}}(\tau,\s)\right].
\ee
To find this generating function,  we use the parameterisation \eqref{Phi_in_out}, which for the canonical momenta yields
\be\label{Pi=zeta}
\Pi_{\text{out}}=\frac{1}{2}\left(\f{\xi''}{\xi'}+\f{\bar\xi''}{\bar\xi'}\right)-\f{\mu}{2}\left(\xi'+\bar\xi'\right)~, \qquad
\Pi_{\text{in}}=\frac{1}{2}\left(\f{\xi''}{\xi'}+\f{\bar\xi''}{\bar\xi'}\right)+\f{\mu}{2}\left(\xi'+\bar\xi'\right)~.
\ee
Similarly to \eqref{th_out-th_in}, the integrand in \eqref{Phi-Phi} can be written as an exact form and one extracts ${\cal G}$ as
\be\label{cal G=}
{\cal G}=\f{\mu}{2}\int_0^{2\pi}\f{\mbox{d}\s}{2\pi}\,\left[\xi'\log\xi'+\bar\xi'\log\bar\xi'-\xi'\log\bar\xi'-\bar\xi'\log\xi'-2\left(\xi'+\bar\xi'\right)\right]~.
\ee
Note that here $\tau$ is fixed, the arguments of $\xi$ and $\bar\xi$ are $x=\tau+\s$ and $\bar x=\tau-\s$, respectively, 
and $'$ denotes the derivative with respect to the argument. Hence, $\cal G$ 
does not split into the sum of the chiral and antichiral parts and it
is $\tau$-dependent.

From \eqref{Phi_in_out} follows
\be\label{xi=Phi}
\xi'\,\bar\xi'=e^{\Phi_+}~, \qquad \mu(\xi'-\bar\xi')=\p_\s\Phi_-\,,
\ee
with $\Phi_{\pm}=\Phi_{\text{in}}\pm\Phi_{\text{out}}$, 
and the generating function \eqref{cal G=} becomes
\be\label{G(Phi,Phi)}
{\cal G}=\int_0^{2\pi}\f{\text{d}\s}{2\pi}\left[\f{1}{2}\p_\s\Phi_-
\log\left(\f{\sqrt{(\p_\s\Phi_-)^{2}+4\m^2\,e^{\Phi_+}}+\p_\s\Phi_-}
{\sqrt{(\p_\s\Phi_-)^{2}+4\m^2\,e^{\Phi_+}}-\p_\s\Phi_-}\right)-
\sqrt{(\p_\s\Phi_-)^{2}+4\m^2\,e^{\Phi_+}}\right].
\ee
This functional defines the semiclassical
$S$-matrix in the $\Phi$-representation. The details of this correspondence are described in the next section. 

Here we introduce another generating function ${\cal F}$, which is related to the semiclassical $S$-matrix in the Fock space.
It is given as a sum of the chiral and antichiral parts ${\cal F}=F+\bar F$, where $F$ is obtained from the differential form
\be\label{F}
\d{ F}=(\tilde q+q)\d p 
-2i\sum_{m>0}\f{1}{m}\left(b^{\phantom{*}}_m\d b_m^*+a_m^*\d a^{\phantom{*}}_m\right),
\ee
and one has a similar antichiral 1-form for $\d\bar F$. 

Thus, $ F$ is treated as a function of the variables $(p,\, b^*_m,\, a_m)$, with $m>0$, and it  satisfies the equations
\be\label{dF}
\f{\p{ F}}{\p p}=\tilde q+q~,\qquad \f{im}{2}\,\f{\p{F}}{\p a_m}=a_m^*~, \qquad \f{im}{2}\,\f{\p{F}}{\p b_m^*}= b_m~.
\ee
The right hand sides can be obtained as functions of $(p,\,b_m^*,\,a_m)$, using the relations 
between the asymptotic fields \eqref{out=in}, or 
\eqref{FF form of T}. We will discuss this point further in Section \ref{semi1}.

It is not difficult to establish the relation between $F$ and $G$. Indeed,   
the 1-forms in \eqref{th_out-th_in} and \eqref{F} are given as a difference between the presymplectic forms of the asymptotic fields. 
The difference $F- G$ is, therefore, a quadratic combination of the Fourier modes. Using the mode expansion \eqref{FF modes}
to evaluate \eqref{th_out-th_in}, one finds 
\be\label{F=G}
F-G=\f 1 2\,  p\,q_+- i\sum_{m>0}\f{1}{m}\left(b^*_m\, b^{\phantom{*}}_m+a_m^*\, a^{\phantom{*}}_m\right)~,
\ee
with $q_+:=\tilde{q}+q$, and from \eqref{G=1} follows 
\be\label{F=modes}
F=p\,q_+ -p-i\sum_{m>0}\f 1 m\left(b_m^*b^{\phantom{*}}_m+a_m^*a^{\phantom{*}}_m
+b_m^*a^{\phantom{*}}_m-a_m^*b^{\phantom{*}}_m\right).
\ee
Here the independent variables are, as stated before, $p$ and $a^{\phantom{*}}_m,b_m^*$ with $m>0$. We will refer to them 
as holomorphic variables. For later use we introduce  
\be\label{defalpha}
\a_n=\begin{cases} a_n & n>0\\b_n & n<0 \end{cases}
\ee
They do not satisfy any reality condition which would relate $\a_n$ to $\a_{-n}$.  

In Appendix \ref{AppB} we will present another realisation of the generating function $G$. 
But in our further discussion it is $F(p,\a)$ and its Legendre transform which play the central role.

\section{Semiclassical S-matrix}\label{semi}

Assuming that the classical relations
\be\label{variation eq}
\f{\d{\cal G}}{\d\Phi_{\text{out}}(\s)}=\f{1}{2\pi}\,\Pi_{\text{out}}(\s)~,\qquad
\f{\d {\cal G}}{\d\Phi_{\text{in}}(\s)}=-\f{1}{2\pi}\,\Pi_{\text{in}}(\s)~,
\ee
are valid at the quantum level with operator ordering  
such that $\hat\Phi_{\text{out}}$ stands to left of $\hat\Phi_{\text{in}}$,  one obtains the equations
\be\ba\label{Pi_in, Pi_out=}
&\f{1}{2\pi}\,\langle\,\Phi_{\text{out}}\,| \hat \Pi_{\text{out}}(\s)\,|\Phi_{\text{in}}\,\rangle
=-i\hbar \f{\d}{\d\Phi_{\text{out}}(\s)}\langle\,\Phi_{\text{out}}\,|\Phi_{\text{in}}\,\rangle
=\langle\,\Phi_{\text{out}}\,|\Phi_{\text{in}}\,\rangle\, \f{\d{\cal G}}{\d\Phi_{\text{out}}(\s)}~,\\[1mm]
&\f{1}{2\pi}\,\langle\,\Phi_{\text{out}}\,| \hat \Pi_{\text{in}}(\s)\,|\Phi_{\text{in}}\,\rangle
=i\hbar \f{\d}{\d\Phi_{\text{in}}(\s)}\langle\,\Phi_{\text{out}}\,|\Phi_{\text{in}}\,\rangle=-
\langle\,\Phi_{\text{out}}\,|\Phi_{\text{in}}\,\rangle\, \f{\d{\cal G}}{\d\Phi_{\text{in}}(\s)}~.
\ea\ee
Up to a constant factor, they are solved by 
\be\label{S(Phi,Phi)}
\langle\,\Phi_{\text{out}}\,|\,\Phi_{\text{in}}\,\rangle\ = e^{\f{i}{\hbar}\,{\cal G}[\Phi_{\text{out}},\Phi_{\text{in}}]}~.
\ee
As the assumptions are generally valid up to ${\cal O}(\hbar)$ terms, this only holds semiclassically.
Due to the complicated form of the functional ${\cal G}$ given in \eqref{G(Phi,Phi)}, eq.\eqref{S(Phi,Phi)} is not convenient 
for further analysis of the transition amplitudes in Fock space.

We now consider the quantum mechanical treatment of the generating function $F$. 
The Fourier-mode operators of the $in$-field satisfy the commutation relations
\be\label{Canonical commutators}
[\hat q, \hat p]=i\hbar~,\quad [\hat a^{\phantom{\dagger}}_m, \hat a_m^\dag]=\tfrac{1}{2}{\hbar m}~,
~~~\hbox{where}~~~~m>0~~~~\hbox{and}~~~  \hat a_m^\dag=\hat a^{\phantom{\dagger}}_{-m}~.
\ee
The $p$-dependent ($p>0$) vacuum state for the chiral $in$-field is defined in the standard way as
\be\label{vacuum state}
\hat p|p, 0\rangle =p|p, 0\rangle~, \qquad \hat a_{m}|p, 0\rangle =0~, \quad \mbox{for}\quad m>0,
\ee
and coherent states are constructed as
\be\label{coherent state}
|p, a\rangle =\exp\left({\f{2}{\hbar}\sum_{m>0} \f{1}{m}\,a^{\phantom{\dagger}}_m\,\hat a_m^\dag}\right)|p, 0\rangle~.
\ee
The $out$-field operators and bra-vectors  $\langle  b^*, \tilde p|$, with $\tilde p<0$,  are defined similarly and one gets
\be\ba\label{Coherent states=}
&\hat a_{m}\,|p, a\rangle =a_m\,|p, a\rangle~, \quad &\hat a_m^\dag\,|p, a\rangle =\tfrac{1}{2}{\hbar\, m}\,\f{\p}{\p a_m}\,|p, a\rangle~, \\[1mm]
&\langle  b^*, \tilde p|\,\hat b_m^\dag=b_m^*\,\langle  b^*, \tilde p|~, \quad &\langle  b^*, \tilde p|\,\hat b_{m}
=\tfrac{1}{2}{\hbar\, m}\,\f{\p}{\p b_m^*}\,\,\langle  b^*, \tilde p|~.
\ea\ee

To analyze the matrix elements $\langle  b^*, p'|p, a\rangle$, we insert the canonical operators and find
\be\label{Eq 1}
\langle  b^*, \tilde p|\hat{\tilde p}+\hat p| p, a\rangle =(\tilde p+p)\langle  b^*,\tilde p|p, a\rangle~,\quad
\langle  b^*, \tilde p|\hat{\tilde q}+\hat q| p, a\rangle =i\hbar (\p_{\tilde{p}}-\p_{p})\langle  b^*,\tilde{p}| p, a\rangle~. 
\ee
If we require $\hat{\tilde{p}}=-\hat p$,
which we know to be satisfied by the classical solutions, 
the left hand side of the first equation vanishes and yields
\be\label{S(p,b*,a)}
\langle  b^*, \tilde p|p, a\rangle=\mathcal{S}(p,b^*\!, a)\,\d(p+\tilde p)~.
\ee
Assuming that the classical relations \eqref{dF} are valid at the quantum level, and that
the operators $\hat b_n^*$ in $F(b^*, p, a)$ stand to the left of the operators $\hat a_{n}$, 
we obtain from \eqref{Coherent states=}-\eqref{S(p,b*,a)}
\be\label{dS(p,b*,a)=}
-i\hbar\, \f{\p \mathcal{S}}{\p p}=\f{\p F}{\p p}\,\mathcal{S}~, \qquad \hbar\, \f{\p \mathcal{S}}{\p a_m}
=i\, \f{\p F}{\p a_m}\,\mathcal{S}~, \qquad
\hbar\, \f{\p \mathcal{S}}{\p b_m^*}=i\, \f{\p F}{\p b_m^*}\,\mathcal{S}~.
\ee
These equations are similar to \eqref{Pi_in, Pi_out=}, and we obtain semiclassically
\be\label{S(p,b*,a)=}
\mathcal{S_0}(p,b^*\!, a)=e^{\f{i}{\hbar}\,F(p,b^*\!,\,a)}~.
\ee

\subsection{Semiclassical amplitudes 1}\label{semi1}

Our goal is to find the generating function $F$ in terms of the holomorphic variables, 
i.e. $F(p,b^*\!,a)$. 
To construct this function from eqs.\eqref{dF}, one has to express their right hand sides as functions 
of $(p,b^*\!, a)$, using the relations between the $in$- and $out$-fields, and then integrate.
The relation between $(q_+,b,a^*)$ and $(p,b^*\!,a)$ is encoded in
\be\label{in=out1}
\phi_{\text{in}}(x)+\phi_{\text{out}}(x)=\log\left[\phi_{\text{in}}'(x)-\phi'_{\text{out}}(x)\right]-\log\m~,
\ee
which follows from \eqref{chiral parts}. This equation contains the full dynamical content of classical  scattering of 
Liouville theory on the cylinder and we can use it to compute $F$, as we will now demonstrate.  The reason for such 
a simple relation to exist is the classical integrability of the theory. 
   
Expanded in Fourier modes it yields
\be\label{in=out2}
q+\tilde{q}+i\sum_{n\neq 0}\f{a_n+b_n}{n}\,e^{-inx}=\log\big(p/\m\big)
+\log\left(1+\sum_{n\neq 0}\f{a_n-b_n}{p}\,e^{-inx}\right)~,
\ee
from which we find the relations
\be\ba\label{q+q=}
&q+\tilde{q}=\log\big(p/\m\big)-\f{1}{p^2}\sum_{m>0}(a_{m}^*-b_{m}^*)(a^{\phantom{*}}_m-b^{\phantom{*}}_m)+\dots~,\\[1mm]
&i\,\f{a_n+b_n}{n}=\f{a_n-b_n}{p}+\dots~.
\ea\ee
The modes $b_m$, $a_m^*$, for $m>0$, and $q+\tilde{q}$ can be obtained from these 
relations in power series of the holomorphic variables $(b^*, a)$ with $p$-dependent coefficients
\be\ba\label{a=b}
&b_m= \f{m-ip}{m+ip}\,a_m+\dots~ , \qquad  \quad a_m^*= \f{m-ip}{m+ip}\,b_m^*+\dots ~, \\[1mm]
&q+\tilde{q}=\log\big(p/\m\big)-\sum_{m>0}\f{4\,b_m^*\,a^{\phantom{*}}_m}{(m+ip)^2}+ \dots ~.
\ea\ee
Integration of \eqref{dF} gives $F(b^*, p, a)=F^{(0)}+F^{(2)}+F^{(3)}+\cdots$, 
where the upper index indicates the total degree of the holomorphic variables. From \eqref{a=b} we then have
\be\label{F0,2}
F^{(0)}=p\left[\log\big(p/\m\big)-1\right]~,\qquad F^{(2)}= -2i\sum_{m>0}\f{1}{m}\,\left(\f{m-i\,p}{m+i\,p}\right)\,b_m^*\,a_m^{\phantom{*}}~.
\ee
The term $F^{(0)}(p)$, given in \eqref{F0,2}, defines the semiclassical reflection 
amplitude
\be\label{R_0}
R_{0}(p)=e^{\f{2i}{\hbar}\,F^{(0)}}~
\ee
of the $p$-dependent vacuum state $|p,\,0\rangle$. 
The total reflection amplitude contains the same contribution from the antichiral part. This 
doubling is taken into account in \eqref{R_0},   
which is common for all transition amplitudes. The higher level states have additional contributions.
For instance, from the expression for $F^{(2)}$ we see that the scattering of the $in$-state $\hat a_m^\dag|p,\,0\rangle$ 
is given by the additional phase factor $\f{m-ip}{m+ip}$.

The calculation of $F^{(3)}$ requires the next order terms in \eqref{q+q=}-\eqref{a=b}, and the continuation of this procedure 
defines $F(p,b^*,a)$ as a power series in holomorphic variables $(b^*, a)$ with $p$-dependent coefficients.

In the following we will present two alternatives to obtain the $F^{(\nu)}$, both of which are more efficient and they can be
generalised to obtain the quantum scattering amplitudes. 

\subsection{Relation to classical one-dimensional field theory}\label{semi2}

We will now use the relation between the $in$- and $out$-fields to construct the Legendre transform of $F$, or rather of
a slightly modified version of $F$, which we will call $\tilde F$; it will be defined momentarily. 
As a consequence of the general relation between semiclassical scattering amplitudes 
and tree-level Feynman diagrams \cite{BB} we expect that the Legendre transform of $\tilde F$, which we will call $S$, is the action 
from which the Feynman rules are derived and the tree-level Feynman diagrams generate $\tilde F$ and, therefore, $F$. 
In the next section we will show that the quantum corrections to the scattering amplitudes are contained in the loop diagrams 
derived from $S$. 

To proceed, we first rewrite eq.\eqref{in=out1}  in the form
\be\label{in=out3}
\phi_{\rm in}'-\phi_{\rm out}'=\mu\,e^{\phi_{\rm in}+\phi_{\rm out}}.
\ee 
We express this in terms of the modes
\be\label{in=out4}
p+{\sum_{n\neq0}}\beta_n\,e^{-inx}=\mu\, e^{q_+}\,e^{i\sum_{n\neq0}{1\over n}\gamma_n\,e^{-inx}},
\ee 
where we have defined the combinations
\be
\gamma_n=a_n+b_n\,,\qquad\beta_n=a_n-b_n\,.
\ee
They are related to $\a_n$, which were defined in \eqref{defalpha} via
\be
\a_n=\tfrac{1}{2}\gamma_n+\tfrac{1}{2}\epsilon(n)\beta_n\,.
\ee 
Here $\epsilon(n)$ is the sign-function, i.e. $\e(n)=+1$ for $n>0$ and $\e(n)=-1$ for $n<0$. 
It will turn out to be convenient to rescale the modes as
\be\label{aJ}
\a_n=-\f{i\epsilon(n)}{ 2}j_n\,,
\ee
with 
\be\label{j}
j_n=i\,\epsilon(n)\gamma_n+i\,\beta_n\,.
\ee
If we define\footnote{Even though we have no compelling a priori reason for this definition, we want to remark that 
the terms which are subtracted from $F$ to obtain $\tilde F$ are those terms in $F$ which survive in the limit $p\to\infty$, 
while keeping all other parameters, including the mode numbers, finite.}
\be\label{Ft}
\tilde F=F-F^{(0)}-i\sum_{n\neq0}{\a_n\a_{-n}\over|n|}=F-p\Big(\ln\frac{p}{\mu}-1\Big)-2\,i\sum_{m>0}{1\over m}b_m^* a_m^{\phantom{*}} \,,
\ee
we find
\be\label{phi_n}
{\p\tilde F\over\p j_{-n}}={1\over n}\g_n\equiv -i\,\varphi_n\,.
\ee
Note that while the $\a_n$ and therefore also the $j_n$ did not 
satisfy any reality condition, $\varphi_{n}^*=\varphi_{-n}$, i.e. they are the Fourier modes of the real field
$\varphi=\phi_{\rm in}+\phi_{\rm out}$. 

If we now use  
\be
\ba
\beta_n&=\mu\,e^{q_+}\int_0^{2\pi}e^{inx}e^{i\sum_{k\neq0}{1\over k}\gamma_ke^{-ikx}}
=\mu\, e^{q_+}\int_0^{2\pi}{dx\over 2\pi} e^{inx}e^{\sum_{k\neq0}\varphi_ke^{-ikx}}\,,
\ea
\ee
which follows from \eqref{in=out4}, we can write \eqref{j} in the form
\be\label{jphi}
j_n=|n|\varphi_n+i\,\mu\,e^{q_+}\!\int_0^{2\pi}{dx\over 2\pi}e^{inx}e^{\sum_{k\neq0}\varphi_ke^{-ikx}}
\mathrel{\mathop=^!}{\p S\over\p\varphi_{-n}}\,,
\ee
where $S$ is defined to be the Legendre transform of $\tilde F$. 

While $j_0$ has not yet been defined, the above relation has an obvious extension to  $n=0$ with 
$j_0=ip$. Here we have used   
\be
p=\mu\,e^{q_+}\int_0^{2\pi}{dx\over 2\pi}e^{i\sum_{k\neq0}{1\over k}\gamma_k e^{-ikx}}\,,
\ee
which is the zero mode of \eqref{in=out4}. It is now also natural to extend \eqref{phi_n} to $n=0$ and to define
\be\label{qtildeplus}
\tilde q_+\equiv\varphi_0={\p\tilde F\over\p p}=q_+-\ln{p\over\mu}\,,
\ee
such that $\mu\,e^{q_+}=p\,e^{\tilde q_+}= p\,e^{\varphi_0}$. 

We now integrate \eqref{jphi} and find 
\be\label{Smodes}
S={1\over2}\sum_{n\in\zz}\varphi_{-n}|n|\varphi_n+i\,p\int_0^{2\pi}{dx\over 2\pi}e^{\sum_{k\in\zz}\varphi_k e^{-ikx}}\,.
\ee
This finally gives a non-local action with exponential potential and an imaginary coupling constant
\be\label{S}
S=\int_0^{2\pi}{dx\over 2\pi}\left(\tfrac{1}{2}\varphi(x)\sqrt{-\p_x^2}\,\varphi(x)+i\,p\,e^{\varphi(x)}\right)\,,
\ee
when written in `position space'.
Note that in this expression $p$ appears as a coupling constant. It is not a dynamical variable. Those are 
$\varphi_n,\,n\in\zz$. 

To recapitulate, we have introduced the conjugated pairs $\{j_n,\varphi_{-n}\}$ and the two functions $\{\tilde F(j),S(\varphi)\}$ which 
are Legendre transformed of each other
\be\label{FS}
S(\varphi)=\langle\varphi\,j\rangle-i\,\tilde F(j)~,
\ee
where 
\be
\langle\varphi\,j\rangle=\int_0^{2\pi}{dx\over 2\pi}\,j(x)\varphi(x)=\sum_{n\in\zz} j_n\,\varphi_{-n}\,.
\ee
In other words,  
\be
e^{\f{i}{\hbar}\tilde F(j)}=\int D\varphi\, e^{-\f{1}{\hbar}(S(\varphi)-\langle\varphi\, j\rangle)}\Big|_{{\d S\over\d\varphi}=j}\,,
\ee
where the functional integral over $\varphi$ is evaluated at the saddle point. In the next section we will include
quantum corrections, i.e. fluctuations around the saddle point. 

As mentioned before, $\tilde F$ is the generating functional for the tree level graphs computed with $S$ or, more 
precisely, due to the relation \eqref{S(p,b*,a)=}, of the connected graphs. We will now compute the lowest order 
$n$-point functions and will then verify them by an explicit calculation using  \eqref{FF form of T}. 

The Feynman rules derived from \eqref{S} are:\footnote{Here we set $\hbar=1$.} 

\begin{enumerate}

\item Draw all connected $\nu$-point tree diagrams with $r$-valent vertices for all $r\geq3$. Each external line 
carries an index $n_i$, $i=1,\dots,\nu$. The labels of the internal lines are dictated by `energy' conservation 
at each vertex. 
 
\item To each $r$-valent vertex assign a factor $-ip$. 

\item To each line, including the external ones, assign a propagator $\f{1}{|n|+ip}$ where $n$ is the `energy' carried by the 
line.  All external lines are ingoing. 
 
\item Sum over all distinct diagrams.  

\end{enumerate}

Note that in rule 3.~we have included the quadratic term of the potential into the propagator. 

We now illustrate these rules.  We expand\footnote{For $\nu\geq3$ the expansions for $F$ and $\tilde F$ coincide.} 
\be\label{Ftexpanded}
\tilde F=\sum_{\nu=2}^\infty\f{1}{\nu!}\,\tilde f^{(\nu)}_{n_1,\dots,n_\nu}\,j_{n_1}\cdots j_{n_\nu}\delta_{n_1+\cdots+n_\nu}~,
\ee
and write the amplitudes for $\nu\geq3$ in the form 
\be
i\tilde f^{(\nu)}_{n_1\dots n_\n}=-ip\,\left(\prod_{j=1}^\n\f{1}{|n_j|+ip}\right)\, U^{(\nu)}_{n_1\dots n_{\n}}~,
\ee
where we have factored out the contributions from the external lines and one power of the coupling constant, i.e. the 
$\nu$-point vertex contribution. We will also use $i\tilde f^{(2)}={1\over|n|+i p}$.
Therefore 
\be
iF^{(2)}=i\sum_{n\neq0} \tfrac{1}{2}\tilde f^{(2)}j_n\,j_{-n}-\sum_{n\neq0}\tfrac{1}{|n|}{\a_n\,\a_{-n}}
=\sum_{n\neq0}\f{1}{|n|}\left(\f{|n|-i\,p}{|n|+i\,p}\right)\a_n\a_{-n}\,,
\ee
which agrees with \eqref{F0,2}. 

We will now evaluate a few tree-level diagrams. Loop diagrams will be considered in Section \ref{quantum}.

The simplest diagram is the three-point function:  there is only the three-point vertex contribution
\be
%\begin{figure}[htb]
\begin{tikzpicture}[scale=.4]
\draw[style=thick] (0,0)--(-1.5,0) node [shift={(-.2,0)}] {$\scriptstyle n_1$};
\draw[style=thick](0,0)--(1,1) node [shift={(0.2,.1)}] {$\scriptstyle n_2$};
\draw[style=thick](0,0)--(1,-1) node [shift={(0.2,-.1)}] {$\scriptstyle n_3$};
\end{tikzpicture}
%\end{figure}
\ee
According to the Feynman rules, it evaluates to
\be
-ip \prod_{i=1}^3\f{1}{ |n_i|+ip}\,,
\ee 
and therefore $U^{(3)}=1$. 

For the four-point function there are four distinct diagrams: a direct four-point vertex and three exchange diagrams
($s,t,u$-channels in the language of particle physics):
\be
%\begin{figure}[htb]
\begin{tikzpicture}[scale=.4]
\draw[style=thick] (0,0)--(-1,1) node [shift={(-.2,.1)}] {$\scriptstyle n_2$};
\draw[style=thick](0,0)--(1,1) node [shift={(0.2,.1)}] {$\scriptstyle n_3$};
\draw[style=thick](0,0)--(-1,-1) node [shift={(-0.2,-.1)}] {$\scriptstyle n_1$};
\draw[style=thick](0,0)--(1,-1) node [shift={(0.2,-.1)}] {$\scriptstyle n_4$};
\draw[style=thick](0,0)--(0,0) node [shift={(1.3,0)}] {$+$};
\end{tikzpicture}
\hspace{.2cm}
\begin{tikzpicture}[scale=.4]
\draw[style=thick] (0,0)--(-1,1) node [shift={(-.2,0)}] {$\scriptstyle n_2$};
\draw[style=thick](0,0)--(-1,-1) node [shift={(-0.2,.-.1)}] {$\scriptstyle n_1$};
\draw[style=thick](0,0)--(2,0) node [shift={(-0.4,.-.2)}] {$\scriptstyle n_1+n_2$};
\draw[style=thick](2,0)--(3,1) node [shift={(0.2,0)}] {$\scriptstyle n_3$};
\draw[style=thick](2,0)--(3,-1) node [shift={(0.2,-.1)}] {$\scriptstyle n_4$};
\draw[style=thick](0,0)--(0,0) node [shift={(2.0,0)}] {$+$};
\end{tikzpicture}
\hspace{.2cm}
\begin{tikzpicture}[scale=.4]
\draw[style=thick] (0,-1)--(-1,-1.5) node [shift={(-.2,0)}] {$\scriptstyle n_1$};
\draw[style=thick](0,-1)--(1,-1.5) node [shift={(0.2,0)}] {$\scriptstyle n_4$};
\draw[style=thick](0,-1)--(0,1) node [shift={(-0.5,.-.35)}] {$\scriptstyle n_1+n_4$};
\draw[style=thick](0,1)--(1,1.5) node [shift={(0.2,0)}] {$\scriptstyle n_3$};
\draw[style=thick](0,1)--(-1,1.5) node [shift={(-0.2,0)}] {$\scriptstyle n_2$};
\draw[style=thick](0,0)--(0,0) node [shift={(1.3,0)}] {$+$};
\end{tikzpicture}
\hspace{.2cm}
\begin{tikzpicture}[scale=.4]
\draw[style=thick] (2,0)--(-1,1) node [shift={(-.2,0)}] {$\scriptstyle n_2$};
\draw[style=thick](0,0)--(-1,-1) node [shift={(-0.2,-.1)}] {$\scriptstyle n_1$};
\draw[style=thick](0,0)--(2,0) node [shift={(-0.4,.-.2)}] {$\scriptstyle n_1+n_3$};
\draw[style=thick](0,0)--(3,1) node [shift={(0.2,0)}] {$\scriptstyle n_3$};
\draw[style=thick](2,0)--(3,-1) node [shift={(0.2,-.1)}] {$\scriptstyle n_4$};
\end{tikzpicture}
%\end{figure}
\ee
The first diagram evaluates to 
\be
-ip\prod_{i=1}^4\f{1}{|n_i|+ip}\,,
\ee
while the second, the $s$-channel diagram, becomes 
\be
(-ip)^2 \left(\prod_{i=1}^4\f{1}{|n_i|+ip}\right)\cdot\f{1}{|n_1+n_2|+ip}\,.
\ee
If we sum over all four diagrams we obtain 
\be\label{U4}
U_{n_1\,n_2\,n_3\,n_4}=1-\f{i\,p}{|n_{1}+n_{2}|+i\,p}-\f{i\,p}{|n_{1}+n_{3}|
+i\,p}-\f{i\,p}{|n_{1}+n_{4}|+i\,p}~.
\ee

\bigskip
As a last example we consider the five-point function. There are now three topologically distinct diagrams:
\be
%\begin{figure}[htb]
\begin{tikzpicture}[scale=.6]
%\draw[style=thick](0,0)--(0:1) node [shift={(.2,0)}] {$\scriptstyle n_3$};
%\draw[style=thick](0,0)--(72:1) node [shift={(.15,.15)}] {$\scriptstyle n_2$};
%\draw[style=thick](0,0)--(144:1) node [shift={(-.2,.2)}] {$\scriptstyle n_1$};
%\draw[style=thick](0,0)--(216:1) node [shift={(-.2,-.2)}] {$\scriptstyle n_5$};
%\draw[style=thick](0,0)--(278:1) node [shift={(.1,-.2)}] {$\scriptstyle n_4$};
\draw[style=thick](0,0)--(0:1) node [shift={(.2,0)}] {$\scriptstyle n_3$};
\draw[style=thick](0,0)--(72:1) node [shift={(.15,.15)}] {$\scriptstyle n_2$};
\draw[style=thick](0,0)--(144:1) node [shift={(-.2,.2)}] {$\scriptstyle n_1$};
\draw[style=thick](0,0)--(216:1) node [shift={(-.2,-.2)}] {$\scriptstyle n_5$};
\draw[style=thick](0,0)--(278:1) node [shift={(.1,-.2)}] {$\scriptstyle n_4$};
\end{tikzpicture}
\hspace{1cm}
\begin{tikzpicture}[scale=.5]
\draw[style=thick] (-1,0)--(0,0) node [shift={(-.7,0)}] {$\scriptstyle n_1$};
\draw[style=thick] (0,0)--(2,0) node [shift={(-.45,-.2)}] {$\scriptstyle n_3+n_4$};
\draw[style=thick](0,0)--(0,1) node [shift={(0,.15)}] {$\scriptstyle n_2$};
\draw[style=thick](0,0)--(0,-1) node [shift={(0,-.15)}] {$\scriptstyle n_5$};
\draw[style=thick](2,0)--(3,1) node [shift={(.2,.2)}] {$\scriptstyle n_3$};
\draw[style=thick](2,0)--(3,-1) node [shift={(.2,-.2)}] {$\scriptstyle n_4$};
\end{tikzpicture}
\hspace{1cm}
\begin{tikzpicture}[scale=.5]
\draw[style=thick] (-1,-1)--(0,0) node [shift={(-.6,-.7)}] {$\scriptstyle n_5$};
\draw[style=thick](-1,1)--(0,0) node [shift={(-.6,.7)}] {$\scriptstyle n_1$};
\draw[style=thick](0,0)--(1.5,0) node [shift={(-.45,-.3)}] {$\scriptstyle n_1+n_5$};
\draw[style=thick](1.5,0)--(3,0) node [shift={(-.2,-.3)}] {$\scriptstyle n_3+n_4$};
\draw[style=thick](3,0)--(4,1) node [shift={(.2,.2)}] {$\scriptstyle n_3$};
\draw[style=thick](3,0)--(4,-1) node [shift={(.2,-.2)}] {$\scriptstyle n_4$};
\draw[style=thick](1.5,0)--(1.5,1) node [shift={(0,.2)}] {$\scriptstyle n_2$};
\end{tikzpicture}
%\end{figure}
\ee

\noindent
For a given `energy' assignment of the external lines, there are $\left({5\atop2}\right)=10$ distinct channels with the topology 
of the second diagram and $\f{1}{2}\left({5\atop2}\right)\left({3\atop2}\right)=15$ for the third diagram. 
The first diagram gives 
\be
-ip \prod_{i=1}^5\f{1}{|n_i|+ip}.
\ee
The second diagram evaluates to 
\be
-ip \left(\prod_{i=1}^5\f{1}{|n_i|+ip}\right)\cdot \f{-ip}{|n_3+n_4|+ip}
\ee
and the third diagram to 
\be
-ip  \left(\prod_{i=1}^5\f{1}{|n_i|+ip}\right)\cdot \f{-ip}{|n_1+n_5|+ip}\cdot\f{-ip}{|n_3+n_4|+ip}\,.
\ee
Summing over all diagrams we find 
\be\label{U5}
U^{(5)}=1+\sum_{i<j}\frac{(-i\,p)}{|n_i+n_j|+i\,p}+\sum_{\{i<j\}\cap\{k<l\}=0}\f{(-i\,p)}{|n_i+n_j|+i\,p}\cdot\f{(-i\,p)}{|n_k+n_l|+i\,p}
\ee
and so on for the higher $U^{(\nu)}$. 

\subsection{Semiclassical amplitudes 2}\label{semi3}

We will now compute $F$ directly. In Section \ref{semi1} we have shown how to use \eqref{in=out1} and the relations \eqref{dF} 
to compute $F$. In this way we have obtained $F^{(0)}$ and $F^{(2)}$ and in principle one can continue to higher orders.
A more efficient algorithm for finding higher order terms of $F$ can be extracted from relation \eqref{FF form of T}, which 
equates the free-field improved energy momentum tensors expressed through the asymptotic 
$in$- and $out$-fields. With the mode-expansions \eqref{FF modes} and
\be\label{Virasoro}
T(x)=\sum_{n\in \zz} L_n\,e^{-inx}~,
\ee
one obtains $L_n$'s in terms of the Fourier modes of the $in$-field
\be\label{L_n}
L_0=\tfrac{1}{4}\,p^2+\sum_{k\neq 0}a_{-k}\,a_k~, \qquad L_n=(p+i\,n)a_n+\sum_{k,l\neq 0}a_k\,a_l\,\d_{k+l,n}~,~~~(n\neq 0)\,.
\ee
The $out$-field mode expansion of $L_n$'s is obtained by replacing $p\mapsto -p$, $a_n\mapsto b_n$. 

Using complex conjugation and only positive indices for the Fourier modes of the asymptotic fields, the equality 
between $L_n$'s of the $in$ and $out$-fields can be written in the form
\be
\ba\label{Ln=Ln}
\sum_{j\geq 1}a_{j}^*\,a^{\phantom{*}}_j&=\sum_{j\geq 1}b_{j}^*\,b^{\phantom{*}}_j~,\\[1mm]
(p+i m)a_m+(p-im)b_m
&=\sum_{j,j'\geq 1}\big[\left(b_jb_{j'}- a_ja_{j'}\right)\d_{j+j',m}+2\left(b_j^{\phantom{*}}b_{j'}^*-a^{\phantom{*}}_ja_{j'}^*\right)\d_{j-j',m}\big],\\ 
\noalign{\vspace{.2cm}}
(p-im)a_{m}^*+(p+im)b_m^*
&=\sum_{j,j'\geq 1}\left[\big(b_j^*b_{j'}^*- a_j^*a_{j'}^*\right)\d_{j+j',m}
+2\left(b_j^*b^{\phantom{*}}_{j'}-a_j^*a^{\phantom{*}}_{j'}\right)\d_{j-j',m}\big]\,,
\ea
\ee
where $m>0$. These relations become equations for $(b,\, a^*)$. They have solutions as 
power series in the holomorphic variables $(b^*,a)$ with $p$-dependent coefficients. 
The function $F(p,b^*,a)$ is then obtained by integration of \eqref{dF}. This leaves, however, $F^{(0)}$ undetermined. 

Alternatively, replacing the variables $(b,\, a^*)$ in \eqref{Ln=Ln}  by the derivatives of $F$, according to \eqref{dF},
we find differential equations directly for the function $F$ which can be solved as power series in the holomorphic variables. 
This turns out to be the more efficient procedure. 
In order to get a covariant structure for scattering amplitudes, 
it is convenient to use instead of the modes $a_m$ and $b^*_m$ for $m>0$, the modes $\a_n$ which we defined in 
\eqref{defalpha} for all integer $n\neq 0$. Doing this we find from \eqref{Ln=Ln} the system of first order 
non-linear differential equations fo $F(p,\a)$
\be\label{eq from L0}
\sum_{k\neq 0} k\,\a_k\f{\p F}{\p \a_k}=0
\ee
and
\be\label{eq from Ln}
\ba
|n|(|n| +ip)\,\f{\p F}{\p \a_{n}}+{2i(|n|-ip)}\,\a_{-n}+i\sum_{k,l\neq 0}\left[\e(k)-\e(l)\right]
|l|\,\a_k\,\f{\p F}{\p \a_{-l}} \,\d_{k+l+n}\\[1mm]
~~~~+\sum_{k,l\neq 0} \left[\e(k)+\e(l)\right]\left(\a_{k}\,\a_{l}+\frac{1}{4}|k|\, |l|\,\f{\p F}{\p \a_{-k}}\,\f{\p F}{\p \a_{-l}}\right)\d_{k+l+n}=0~,
\ea
\ee
where $n$ is a non-zero integer. 

A monomial $\a_{n_1}\cdots \a_{n_\n}$ solves the linear equation \eqref{eq from L0} if  $n_1+\cdots+n_\n=0$
and the function $F$ is then represented as a power series; cf. \eqref{Ft} and \eqref{Ftexpanded}.  

At this point it turns out to be convenient to introduce $\tilde F$ and the $j_n$ of Section \ref{semi2}. They lead to the 
simpler equation for $\tilde F$: 
\be\label{eqtildeF}
\ba
in\big(|n|+ip\big)\f{\p\tilde F}{\p j_n}-nj_{-n}-i\sum_{k,l\neq0}l j_k\f{\p\tilde F}{\p j_{-l}}\delta_{k+l+n}
-\sum_{k,l\neq0}|k|l\f{\p\tilde F}{\p j_{-k}}\f{\p\tilde F}{\p j_{-l}}\delta_{k+l+n}=0~.
\ea
\ee
We insert the Ansatz \eqref{Ftexpanded} and solve \eqref{eqtildeF} order by order in the $\a_n$ and determine 
the amplitudes $f^{(\nu)}$ recursively.   

At the linear order, we find 
\be
i\,\tilde f^{(2)}={1\over|n|+i\,p}
\ee
and at the quadratic order 
\be
i\,\tilde f^{(3)}=-i\,p\cdot\prod_{i=1}^3{1\over|n_i|+i\,p}\,.
\ee
Both are in agreement with the results presented in Section \ref{semi2}.
The first non-trivial case comes from solving \eqref{eqtildeF} at cubic order. After a somewhat tedious calculation 
one finds again complete agreement with the Feynman diagram calculation in Section \ref{semi2}, i.e. with \eqref{U4}. 

Since the relation \eqref{FF form of T}, which we exploited to obtain the differential equation obeyed by $F$, follows  
immediately from  \eqref{in=out1} by differentiating w.r.t. $x$, it should be clear that we can also find $S$ starting from 
\eqref{eqtildeF}. Indeed, using the relations
\be\label{jvarphi}
j_n=\f{\p S}{\p \varphi_{-n}}\,,\qquad
\varphi_n=i\,\f{\p\tilde F}{\p j_{-n}}\qquad\qquad(n\neq0)
\ee
in \eqref{eqtildeF}, it becomes
\be\label{eqS}
n\big(|n|+ip\big)\varphi_{-n}-n\f{\p S}{\p\varphi_n}-\sum_{k,l\neq0}l\,\varphi_l\,\f{\p S}{\p\varphi_{-k}}\,\delta_{k+l+n}
+\sum_{k,l\neq0}|k|l\,\varphi_k\,\varphi_l\,\delta_{k+l+n}=0\,.
\ee
The simplification effected by passing from $F$ to the shifted $\tilde F$ is that the quadratic terms in the modes have 
cancelled in \eqref{eqtildeF} and the functional differential equation satisfied by $S$ is therefore linear. 
If we make the Ansatz
\be\label{AnsatzS}
S=\f{1}{2}\sum_{n\in\zz} |n|\varphi_{-n}\varphi_n+V(\varphi)\,,
\ee
we obtain a simple equation for $V$: 
\be\label{eqV}
-i\,p\,n\,\varphi_{-n}+n\f{\p V}{\p\varphi_n}+\sum_{k,l\neq0}l\varphi_l\f{\p V}{\p\varphi_{-k}}\delta_{k+l+n}=0\,.
\ee
Here a zero mode for $\varphi$ is not included. But we can define  
$\varphi_0$ as the variable conjugate to $j_0={\p V\over\p\varphi_0}$. If we identify the latter with $ip$, 
we find $\varphi_0=i{\p\tilde F\over\p j_0}=\tilde q_+$, as in the discussion in Section \ref{semi1}. 
The equation for $V$ is then solved by $V(\varphi)=i\,p\,e^{\varphi}$.  The constant term in $V$ stays undetermined 
by eq.\eqref{eqV}, but it can be chosen arbitrarily, as it does not affect the amplitudes. Here it is chosen as $ip$.

\section{Quantum $S$-matrix}\label{quantum}

After having computed the semiclassical scattering amplitudes, we now turn to the quantum corrected 
amplitudes. We compute them in two alternative ways. We first use the classical action $S$, from which, as we have just shown,  
the semiclassical amplitudes can be obtained as tree-level Feynman diagrams, to compute loop amplitudes. 
We then generalise the procedure of Section \ref{semi3} by equating the quantum-corrected energy-moment tensors for 
the asymptotic $in$- and $out$-fields. We then compare and find agreement. 
   
The second method is computationally superior and it  
produces results which are valid to all orders in $\hbar$, in agreement with an 
old result of Zamolodchikov and Zamolodchikov \cite{Zamolodchikov:1995aa}.  
We will see that the simplicity of the equation satisfied by 
$\tilde F$ is destroyed and we have not been able to use it to get the full quantum effective action, i.e. the generating 
functional for the quantum scattering amplitudes, by integrating this equation, as we did in Section \ref{semi3}.  

\subsection{Quantum amplitudes 1}\label{quantum1}

The starting point is the action \eqref{Smodes}. The Feynman rules are the same as before, but now we allow loops 
with the only restriction that self-contractions of two lines emanating from the same vertex are not allowed. This 
corresponds to `normal ordering' of the potential. It eliminates all possible divergences and therefore no further 
renormalisation is necessary.\footnote{We have not proven this statement but the overall degree of divergence of 
every diagram is clearly negative.}
We also have to take into account symmetry factors, 
just as in conventional scalar field theory. We now evaluate a few one-loop diagrams. 

For the one-loop contribution of the two-point function there is a single diagram\footnote{Here we use the notation 
$f^{(n,\ell)}$ for the $\ell$-loop contribution for the $n$-point function.} 
\be
%\begin{figure}[htb]
i \tilde f^{(2,1)}_{-n\,n}=\raisebox{-.3cm}{
\begin{tikzpicture}[scale=.4]
\draw[style=thick] (0,0) circle (1);
\draw[style=thick](-2,0)--(-1,0)  node [shift={(-0.7,0)}] {$\scriptstyle n$};
\draw[style=thick](1,0)--(2,0) node [shift={(0.5,0)}] {$\scriptstyle -n$};
\end{tikzpicture}}
%\end{figure}
~~=~~\tfrac{1}{2}(-ip)^2\,\Delta(n)^2\sum_{m=-\infty}^\infty \Delta(m)\Delta(m+n)\,.
\ee
We have defined the propagator
\be
\Delta(n)=\f{1}{|n|\!+\!ip}\,, 
\ee
and we have included a factor $-ip$ for each vertex; $1/2$ is the appropriate symmetry factor.
The (convergent) sum can be computed for general $n$ in terms of the digamma function $\psi$, 
but the expression is not very illuminating and we will not write it. 
A few special cases are 
\be\label{f2quant}
\tilde f^{(2,1)}_{-1\,1}=i\,p\,\Delta(1)^2\,,\quad 
\tilde f^{(2,1)}_{-2\,2}=-\tfrac{1}{2}i\,p\,(2 p^2-4\,i\,p-1)\Delta(1)^2 \Delta(2)^2\,.
\ee

Next we look at one-loop diagrams with three external legs. There are three possible topologies 
\be
\ba
%\begin{figure}[htb]
&\hspace*{.5cm}\begin{tikzpicture}[scale=.4]
\draw[style=thick] (0,0) circle (1) ;
\draw[style=thick](-2,0)--(-1,0) node [shift={(-.6,0)}] {$\scriptstyle n_1$};
\draw[style=thick](1,0)--(2,1) node [shift={(0.3,0.1)}] {$\scriptstyle n_2$};
\draw[style=thick](1,0)--(2,-1) node [shift={(0.3,-.1)}] {$\scriptstyle n_3$};
\draw[style=thick] (11,0) circle (1) node [shift={(-1.1,0)}] {$\scriptstyle n_1$};
\draw[style=thick](9,0)--(10,0);
\draw[style=thick](12,0)--(13,0);
\draw[style=thick](13,0)--(14,1) node [shift={(0.3,.1)}] {$\scriptstyle n_2$};
\draw[style=thick](13,0)--(14,-1) node [shift={(0.3,-.1)}] {$\scriptstyle n_3$};
\draw[style=thick] (22,0) circle (1);
\draw[style=thick](20,0)--(21,0) node [shift={(-.6,0)}] {$\scriptstyle n_1$};
\draw[style=thick](22.9,.5)--(24,1)  node [shift={(0.3,.1)}] {$\scriptstyle n_2$};
\draw[style=thick](22.9,-.5)--(24,-1) node [shift={(0.3,-.1)}] {$\scriptstyle n_3$};
\end{tikzpicture}
%\end{figure}
\\
&~~~~~\scriptstyle{D_1^{(3,1)}(n_1,n_2,n_3)} \hspace{2.2cm} D_2^{(3,1)}(n_1,n_2,n_3) \hspace{2.2cm} D_3^{(3,1)}(n_1,n_2,n_3) 
\ea
\ee
and the amplitude is 
\be
\ba
&i f^{(3,1)}_{n_1\,n_2\,n_3}=\tfrac{1}{2}\big(D_1^{(3,1)}(n_1,n_2,n_3) +D_1^{(3,1)}(n_2,n_3,n_1) +D_1^{(3,1)}(n_3,n_1,n_2)\\[3mm]
&+ D_2^{(3,1)}(n_1,n_2,n_3) +D_2^{(3,1)}(n_2,n_3,n_1) +D_2^{(3,1)}(n_3,n_1,n_2)\big)+D_3^{(3,1)}(n_1,n_2,n_3)\,. 
\ea
\ee
All lines are ingoing with $n_1+n_2+n_3=0$. 
Using the Feynman rules it is straightforward to evaluate the diagrams; for example
\be
\ba
\hspace*{.5cm}
\raisebox{-.5cm}{\begin{tikzpicture}[scale=.3]
\draw[style=thick] (22,0) circle (1);
\draw[style=thick](20,0)--(21,0) node [shift={(-.6,0)}] {$\scriptstyle n_1$};
\draw[style=thick](22.9,.5)--(24,1)  node [shift={(0.3,.1)}] {$\scriptstyle n_2$};
\draw[style=thick](22.9,-.5)--(24,-1) node [shift={(0.3,-.1)}] {$\scriptstyle n_3$};
%\draw[style=thick] (0,0) circle (1) ;
%\draw[style=thick](-2,0)--(-1,0) node [shift={(-.6,0)}] {$\scriptstyle n_1$};
%\draw[style=thick](1,0)--(2,1) node [shift={(0.3,0)}] {$\scriptstyle n_2$};
%\draw[style=thick](1,0)--(2,0) node [shift={(0.3,0)}] {$\scriptstyle n_3$};
%\draw[style=thick](1,0)--(2,-1) node [shift={(0.3,0)}] {$\scriptstyle n_3$};
\end{tikzpicture}}
&=(-i p)^3 \,\prod_{i=1}^3 \Delta(n_i)\sum_{n=-\infty}^\infty \Delta(n)\Delta(n+n_1)\Delta(n+n_1+n_2)\,.
\ea
\ee
The sum can be performed for any given $n_1, n_2,n_3$ with $\sum n_i=0$.\footnote{e.g. with Mathematica. One can also work 
out a general expression for generic $n_i$ in terms of Polygamma functions, but it is rather lengthy and does not seem to be 
particularly useful.} 
Some examples of the complete amplitudes are
\be
\ba
i \tilde f^{(3,1)}_{-1\,1\,1}&=i p \big(p^2+i p+5\big)\Delta(1)^3 \Delta(2)^2\,,\\[2mm]
i \tilde f^{(3,1)}_{-3\,1\,2}&=-i p\big(p^4-i p^3+14 p^2-29ip-11\big)\Delta(1)^3 \Delta(2)^2 \Delta(3)^2\,.
\ea
\ee

\medskip

For four external lines there are the following ten topologies:  
\be
%\begin{figure}[htb]
%\hspace*{.5cm}
\begin{tikzpicture}[scale=.24]
\draw[style=thick] (0,0) circle (1) ;
\draw[style=thick](-2,0)--(-1,0);
\draw[style=thick](1,0)--(2,1);
\draw[style=thick](1,0)--(2,0);
\draw[style=thick](1,0)--(2,-1);
\hspace{.1cm}
\draw[style=thick] (5,0) circle (1);
\draw[style=thick](3,0)--(4,0);
\draw[style=thick](6,0)--(7,0);
\draw[style=thick](7,0)--(8,1);
\draw[style=thick](7,0)--(8,0);
\draw[style=thick](7,0)--(8,-1);
\hspace{.1cm}
\draw[style=thick] (11,0) circle (1);
\draw[style=thick](9,1)--(10,0);
\draw[style=thick](9,-1)--(10,0);
\draw[style=thick](11.9,.5)--(13,1);
\draw[style=thick](11.9,-.5)--(13,-1);
\hspace{.1cm}
\draw[style=thick] (17,0) circle (1);
\draw[style=thick](14,1)--(15,0);
\draw[style=thick](14,-1)--(15,0);
\draw[style=thick](15,0)--(16,0);
\draw[style=thick](17.9,.5)--(19,1);
\draw[style=thick](17.9,-.5)--(19,-1);
\hspace{.1cm}
\draw[style=thick] (23,0) circle (1);
\draw[style=thick](20,1)--(21,0);
\draw[style=thick](20,-1)--(21,0);
\draw[style=thick](21,0)--(22,0);
\draw[style=thick](24,0)--(25,0); 
\draw[style=thick](25,0)--(26,1);
\draw[style=thick](25,0)--(26,-1);
\hspace{-.1cm}
\draw[style=thick] (30,0) circle (1);
\draw[style=thick](28,1)--(29,0);
\draw[style=thick](28,-1)--(29,0);
\draw[style=thick](31,0)--(32,0); 
\draw[style=thick](32,0)--(33,1);
\draw[style=thick](32,0)--(33,-1);
\hspace{.1cm}
\draw[style=thick] (36,0) circle (1);
\draw[style=thick](34,1)--(35,0);
\draw[style=thick](34,-1)--(35,0);
\draw[style=thick](37,0)--(38,1);
\draw[style=thick](37,0)--(38,-1);
\draw[style=thick] (41,0) circle (1);
\draw[style=thick](39,0)--(40,0);
\draw[style=thick](42,0)--(43,1);
\draw[style=thick](42,0)--(43,0); 
\draw[style=thick](43,0)--(44,1);
\draw[style=thick](43,0)--(44,-1);
\hspace{-.2cm}
\draw[style=thick] (48,0) circle (1);
\draw[style=thick](46,0)--(47,0);
\draw[style=thick](49.5,0)--(49.5,1);
\draw[style=thick](49,0)--(50,0); 
\draw[style=thick](50,0)--(51,1);
\draw[style=thick](50,0)--(51,-1);
\hspace{.1cm}
\draw[style=thick] (54,0) circle (1);
\draw[style=thick](52,1)--(53,.3);
\draw[style=thick](52,-1)--(53,-.3);
\draw[style=thick](55,.3)--(56,1);
\draw[style=thick](55,-.3)--(56,-1);
\end{tikzpicture}
\ee
Using the Feynman rules it is again straightforward to evaluate the diagrams and to add them.  
%
%for example
%\be
%\ba
%\hspace*{.5cm}
%\raisebox{-.5cm}{\begin{tikzpicture}[scale=.3]
%\draw[style=thick] (0,0) circle (1) ;
%\draw[style=thick](-2,0)--(-1,0) node [shift={(-.6,0)}] {$\scriptstyle n_1$};
%\draw[style=thick](1,0)--(2,1) node [shift={(0.3,0)}] {$\scriptstyle n_2$};
%\draw[style=thick](1,0)--(2,0) node [shift={(0.3,0)}] {$\scriptstyle n_3$};
%\draw[style=thick](1,0)--(2,-1) node [shift={(0.3,0)}] {$\scriptstyle n_4$};
%\end{tikzpicture}}
%&=\tfrac{1}{2}(-i p)^2 \,\prod_{i=1}^4 \Delta(n_i)\sum_{n=-\infty}^\infty \Delta(n)\Delta(n+n_1)\,.
%\\
%&\equiv \tilde D_4^{(4,1)}(n_1,n_2,n_3,n_4) \prod_{i=1}^4P(n_i) 
%\ea
%\ee
%The sums can be performed with Mathematica for any given $n_1, n_2,n_3,n_4$ with $\sum n_i=0$.\footnote{One can also work 
%out a general expression for generic $n_i$ in terms of Polygamma functions, but they are rather lengthy and do not seem to be 
%particularly useful.} 
A few examples for complete one-loop four-point amplitudes are
\be\label{f4quant}
\ba
i \tilde f^{(4,1)}_{-1\,-1\,1\,1}&={-2ip(p^2+2ip+7)}\Delta(2)^2\Delta(1)^4\,,\\[2mm]
%i f^{(4,1)}(-2,-1,1,2)&=-p(564i-1146 p+137 i p^2-1200 p^3-733 i p^4+146 p^5+2 p^7)P(1)^2 P(2)^2 P(3)^2 P(4)
i \tilde f^{(4,1)}_{-3\,1\,1\,1}&=2ip (p^4+4 i p^3+30 p^2-20 i p+29)\Delta(1)^4\Delta(2)^2\Delta(3)^2\,,\\[2mm]
i \tilde f^{(4,1)}_{-2\,-1\,1\,2}&=2ip (p^4+2 i p^3+25 p^2-30 i p+6)\Delta(1)^3 \Delta(2)^2 \Delta(3)^2\,.
\ea
\ee
It is easy to work out others, but the order of the polynomial in $p$ in the numerator grows fast with the occupation numbers
of the external lines.  

In principle one can go higher in the loop expansion, but the number of diagrams grows quickly and the multiple sums 
can no longer be performed in closed form. For instance, the following diagrams contribute to the two-loop corrections of the propagator

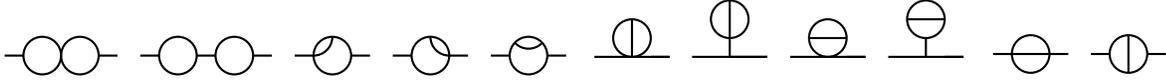
\begin{figure}[htb]
\subfloat{
\begin{tikzpicture}[scale=.25]
\draw[style=thick] (0,0) circle (1);
\draw[style=thick](-2,0)--(-1,0);
\draw[style=thick](3,0)--(4,0);
\draw[style=thick](2,0) circle (1);
\end{tikzpicture}
}
\subfloat{
\begin{tikzpicture}[scale=.25]
\draw[style=thick] (0,0) circle (1);
\draw[style=thick](-2,0)--(-1,0);
\draw[style=thick](1,0)--(2,0);
\draw[style=thick](3,0) circle (1);
\draw[style=thick](4,0)--(5,0);
\end{tikzpicture}
}
\subfloat{
\begin{tikzpicture}[scale=.25]
\draw[style=thick] (0,0) circle (1);
\draw[style=thick](-2,0)--(-1,0);
\draw[style=thick](1,0)--(2,0);
\draw [style=thick] (0,1) arc [radius=1, start angle=0, end angle= -90];
\end{tikzpicture}
}
\subfloat{
\begin{tikzpicture}[scale=.25]
\draw[style=thick] (0,0) circle (1);
\draw[style=thick](-2,0)--(-1,0);
\draw[style=thick](1,0)--(2,0);
\draw [style=thick] (1,0) arc [radius=1, start angle=270, end angle= 180];
\end{tikzpicture}
}
\subfloat{
\begin{tikzpicture}[scale=.25]
\draw[style=thick] (0,0) circle (1);
\draw[style=thick](-2,0)--(-1,0);
\draw[style=thick](1,0)--(2,0);
\draw [style=thick] (0.71,0.66) arc [radius=1, start angle=315, end angle= 225];
\end{tikzpicture}
}
\\[-1.4cm]
\hspace*{7.7cm}
\subfloat{
\begin{tikzpicture}[scale=.25]
\draw[style=thick] (0,1) circle (1);
\draw[style=thick](-2,0)--(2,0);
\draw[style=thick](1,0)--(2,0);
\draw[style=thick](0,0)--(0,2);
\end{tikzpicture}
}
\subfloat{
\begin{tikzpicture}[scale=.25]
\draw[style=thick] (0,2) circle (1);
\draw[style=thick](-2,0)--(2,0);
\draw[style=thick](0,0)--(2,0);
\draw[style=thick](0,0)--(0,3);
\end{tikzpicture}
}
\subfloat{
\begin{tikzpicture}[scale=.25]
\draw[style=thick] (0,1) circle (1);
\draw[style=thick](-2,0)--(2,0);
\draw[style=thick](1,0)--(2,0);
\draw[style=thick](-1,1)--(1,1);
\end{tikzpicture}
}
\subfloat{
\begin{tikzpicture}[scale=.25]
\draw[style=thick] (0,2) circle (1);
\draw[style=thick](-2,0)--(2,0);
\draw[style=thick](1,0)--(2,0);
\draw[style=thick](-1,2)--(1,2);
\draw[style=thick](0,0) -- (0,1);
\end{tikzpicture}
}
\\[-0.7cm]
\hspace*{13cm}
\subfloat{
\begin{tikzpicture}[scale=.25]
\draw[style=thick] (0,0) circle (1);
\draw[style=thick](-2,0)--(2,0);
\end{tikzpicture}
}
\subfloat{
\begin{tikzpicture}[scale=.25]
\draw[style=thick] (0,0) circle (1);
\draw[style=thick](-2,0)--(-1,0);
\draw[style=thick](1,0)--(2,0);
\draw[style=thick](0,-1)--(0,1);
\end{tikzpicture}
}
\end{figure}
\noindent
The double sums seem hard to do analytically, 
but one can do them numerically for some fixed values of $p$ (they converge rather slowly). 
This and the one-loop results can be compared  
with the  expressions obtained by the second method that we have alluded to and to which we now turn.  

\subsection{Quantum amplitudes 2}\label{quantum2}

A consistent quantization of Liouville theory leads to a deformation of the improvement term in the stress tensor  
\cite{Braaten:1982yn, Jorjadze:2001nx}.
The quantum Virasoro generators in terms of the $in$-field variables read
(c.f. \eqref{L_n})
\be\ba\label{Q-Ln}
&\hat L_0=\f{1}{4}(\hat p^2+\eta^2)+2\sum_{j\geq 1}\hat a_j^\dag\,\hat a^{\phantom{\dag}}_j~, \\[2mm]
&\hat L_m=(\hat p+i\,m\,\eta)\hat a_m+\sum_{j, j'\geq 1}\hat a_j\,\hat a_{j'}\,\d_{j+j',m} +
2\sum_{j>0}\hat a_{j}^\dag\,\hat a^{\phantom{\dag}}_{m+j}~,\\[2mm]
%\label{Q-L-n}
&\hat L_{-m}=(\hat p-i\,m\,\eta)\hat a_{m}^\dag+\sum_{j, j'\geq 1}\hat a_{j}^\dag\,\hat a_{j'}^\dag\,\d^{\phantom{\dag}}_{j+j',m} 
+2\sum_{j\geq 1}\hat a_{m+j}^\dag\,\hat a^{\phantom{\dag}}_j~,
\ea\ee
where $m\geq 1$ and $\eta=1+\hbar$. 
The same generators in terms of the $out$-field variables are obtained by the replacements $p \mapsto -p$, $\hat a_j\mapsto \hat b_j$ 
and $\hat a_j^\dag \mapsto \hat b_j^\dag$.

The zero mode operators fix the transition amplitude between the $in$ and $out$ coherent states \eqref{coherent state} 
as in \eqref{S(p,b*,a)}, where the
freedom is given by the function ${\mathcal S}(p,b^*\!,a)$.
Inserting the Virasoro generators between the coherent states 
$\langle\tilde p, b^*| \hat L_n|p, a\rangle$ and using the relations \eqref{Coherent states=},
one obtains equations for the function ${\mathcal S}(p,b^*\!,a)$.

To get a covariant structure of the scattering amplitudes we again combine the positive and negative indices of $L_n$'s, 
as we did in the semiclassical treatment. We then obtain the following equations
\be\label{Q-eq from L0}
\sum_{k\neq 0} k\,\a_k\,\f{\p \mathcal{S}}{\p \a_k}=0~,
\ee
\be\ba\label{Q-eq from Ln=}
\hbar |n|\,(p-i|n|\eta )\f{\p \mathcal{S}}{\p \a_{n}}+2(p+i|n|\eta)\a_{-n}\,\mathcal{S}+
{\hbar}\sum_{k, l\neq 0}\left[\e(k)-\e(l)\right]|l|\, \a_k\,\f{\p \mathcal{S}}{\p \a_{-l}}\,\d_{k+l+n}\\[2mm]
+\sum_{k, l\neq 0} \left[\e(k)+\e(l)\right]\left(\a_{k}\,\a_{l}\,\mathcal{S}-\hbar^2\,\f{|k|\,|l|}{4} 
\f{\p^2 \mathcal{S}}{\p \a_{-k}\,\p \a_{-l}}\right)\d_{k+l+n}=0~.
\ea\ee

We represent  $\mathcal{S}$ in the form\footnote{$F_q$ starts at ${\cal O}(\alpha^2)$. If one includes both chiral sectors, this becomes
${\cal S}=R(p)e^{\f{i}{\hbar}(F_q+\bar F_q)}$; see also the discussion in the Conclusions.}
\be\label{S=}
\mathcal{S}=R(p)\, e^{\f{i}{\hbar} F_q}~,
\ee
where $R(p)$
is interpreted as the reflection amplitude, which is known as the 2-point function of 
Liouville theory \cite{Dorn:1994xn, Zamolodchikov:1995aa}
\be\label{Rp}
R(p)=-\left(\m^2\frac{\sin(\pi\hbar)}{\pi\hbar}\G^2(\hbar)\right)
^{-\frac{ip}{\hbar}}\,\frac{\G(ip/\hbar)}{\G(-ip/\hbar)}\,
\frac{\,\G(ip)}{\,\G(-ip)}~.
\ee
It reduces to \eqref{R_0} in the $\hbar\to0$ limit. 
This factor is canceled in \eqref{Q-eq from L0}-\eqref{Q-eq from Ln=} and $F_q$ satisfies
\be\ba\label{Q-eq for Fq}
{|n|}(|n|\eta +ip)\f{\p F_q}{\p \a_{n}}+2i(|n|\eta-ip)\a_{-n}+
i\sum_{k, l\neq 0} \left[\e(k)-\e(l)\right]|l|\, \a_k\,\f{\p F_q}{\p \a_{-l}}\,\d_{k+l+n}\\[2mm]
+\sum_{k, l\neq 0} \left[\e(k)+\e(l)\right]\left[\a_{k}\,\a_{l}+\f{|k|\,|l|}{4}\left(\f{\p F_q}{\p \a_{-k}}\,\f{\p F_q}{\p \a_{-l}}
-i\hbar \f{\p^2 F_q}{\p \a_{-k}\,\p \a_{-l}}\right)\right]\d_{k+l+n}=0~,
\ea\ee
which is the quantum analogue of \eqref{eq from Ln}.
Planck's constant appears through $\eta$ and in the last term. It is this last term which makes the solution of this 
equation considerably more difficult than in the semiclassical case. 

If we again introduce $\tilde F$ as 
\be\label{Fqtilde}
\tilde F_q=F_q-i\sum_{n\neq0}{\a_n\a_{-n}\over|n|} 
\ee
and perform the change of variables  \eqref{aJ}, as we did in the semiclassical discussion,
we arrive at the equation 
\be\label{eqFredj}
\ba
&i\,n\big(|n|\eta+ip\big)\f{\p\tilde F_q}{\p j_n}-n\,\eta\,j_{-n}-i\sum_{k,l\neq0}l\,j_{k}\f{\p\tilde F_q}{\p j_{-l}}\delta_{k+l+n}\\[2mm]
&\hspace{3cm}
-\sum_{k,l\neq0}|k|l\left(\f{\p\tilde F_q}{\p j_{-k}}\f{\p\tilde F_q}{\p j_{-l}}-i\,\hbar\f{\p^2\tilde F_q}{\p j_{-k}\,\p j_{-l}}\right)\delta_{k+l+n}
=0\,.
\ea
\ee
As before, we make a series Ansatz for $\tilde F$, insert it into (\ref{eqFredj}) and solve for the coefficients. 
However,  in contrast to the semiclassical case, i.e. as a consequence of the second derivative term in 
the quantum equation, we can no longer solve for the $\tilde f^q_{n_1\dots n_\nu}$ for generic mode numbers $n_i$. 
But we can still solve them recursively and it is clear that they are rational functions of $p$ and $\hbar$. 

For small $\nu$ and small mode numbers one can easily find explicit expressions, which are valid to 
all orders in $\hbar$.  A few examples are\footnote{The results at levels one and  two 
agree with \cite{Zamolodchikov:1995aa}.} 
\be\label{fqs}
\ba
&i\tilde f^{(2,q)}_{-1\,1}=\f{1+\hbar}{1+\hbar+ip}\,,\qquad
i \tilde f^{(2,q)}_{-2\,2}=\f{(1 + \hbar) \big(2(1+i p)^2+(5+4ip)\hbar+2\hbar^2\big)}{ 2 (1+\hbar+i p) (2 + \hbar+i p) (1+2 \hbar+i p)}\,,\\[2mm]
&i\tilde f^{(3,q)}_{-2\,1\,1}=-\f{i(1+\hbar) p}{(1+\hbar+i p) (2+\hbar + i p) (1+2 \hbar + i p)}\,,\\[2mm]
&i\tilde f^{(3,q)}_{-3\,1\,2}=-\f{i(1+\hbar) p \big((1+ip)^2+2(2+ip)\hbar+\hbar^2\big)}{(1+\hbar+i p) (2+\hbar+i p) (3+\hbar+i p) 
(1+2 \hbar+i p) (1+3 \hbar+i p)}\,,\\[2mm]
&i\tilde f^{(4,q)}_{-1\,-1\,1\,1}=\f{2 i (1+\hbar) p}{(1 + \hbar+i p)^2 (2+\hbar+i p) (1+2 \hbar+i p)}\,,\\[2mm]
&i\tilde f^{(4,q)}_{-2\,-1\,1\,2}=\f{2 i (1+\hbar)p(1+i p) (\hbar+i p)}{(1+\hbar+i p)^2(2+\hbar+i p)
(3+\hbar+i p) (1+2 \hbar+i p) (1+3 \hbar+i p)}\,,\\[2mm]
&i\tilde f^{(4,q)}_{-3\,1\,1\,1}=-\f{2 i (1+\hbar) p(1+\hbar-i p)}{(1+\hbar+i p) (2+\hbar+i p) (3+\hbar+i p) (1+2\hbar+i p) (1+3 \hbar+i p)}\,.
\ea
\ee
The expressions with higher mode numbers quickly become too long to display. We have also restricted to 
at most four oscillator excitations because they are needed for the comparison with the results obtained from the loop calculations. 
Indeed, expanding to first order in $\hbar$, we find agreement with \eqref{f2quant} and \eqref{f4quant}. We also found agreement 
between the ${\cal O}(\hbar^2)$ terms in $\tilde f^{(2,q)}_{-1\,1}$ and $\tilde f^{(2,q)}_{-2\,2}$ and the numerical results 
of the two-loop amplitudes. 

\section{Conclusions}

Our aim in this paper has been the computation of the $S$-matrix of Liouville theory on a cylinder. 
The analysis of the relation between the asymptotic fields shows that this operator can be represented in the form 
\be\label{S-matrix form}
\hat{\cal S}=\hat{\cal P}\, R(p)\, {\cal S}_p(\hat a_n)\,{\cal S}_p(\hat{\bar a}_n)~,
\ee
where $\hat{\cal P}$ is the parity operator that reflects the zero modes of the $in$-field
\be\label{Parity op}
\hat{\cal P}\, \hat q_{\text{in}} \hat{\cal P}=-\hat q_{\text{in}}~, \qquad 
\hat{\cal P}\, \hat p_{\text{in}} \hat{\cal P}=-\hat p_{\text{in}}~,
\ee
$R(p)$ is the reflection amplitude \eqref{Rp}, ${\cal S}_p(\hat a_n)$ depends only on the creation-annihilation operators of the chiral 
$in$-field, ${\cal S}_p(\hat{\bar a}_n)$ is its antichiral counterpart and
the matrix elements of ${\cal S}_p(\hat a_n)$ in the basis of the coherent states \eqref{coherent state} are defined by the function 
of holomorphic variables ${\cal S}(p,b^*\!,a)$ introduced in \eqref{S(p,b*,a)}.
Since the reflection amplitude of Liouville theory is known, the computation of the $S$-matrix is reduced to the 
analysis of the chiral sector only. 

The relevance of the computation of the chiral part of the $S$-matrix, given by ${\cal S}_p(\hat a_n)$, goes beyond Liouville theory 
on the cylinder \cite{Braaten:1982yn, Otto:1985eb} and extends to the theory on the strip \cite{Gervais:1983ry}, 
which has besides scattering solutions, also bound states \cite{Fateev:2000ik, Ponsot:2001ng}.
In particular, the asymptotic $in$-field of Liouville theory on the strip with Neumann boundary conditions has only one set of Fourier 
modes $a_n$ and the $S$-matrix is represented by \cite{Dorn:2008sw}
\be\label{S-matrix form 1}
\hat{\cal S}=\hat{\cal P}\, R_b(p)\, {\cal S}_p(\hat a_n),
\ee
where $\hat{\cal P}$ is again the parity operator \eqref{Parity op}, $R_b(p)$ is the reflection amplitude of the boundary theory and 
$S_p(\hat a_n)$ is the same operator as in \eqref{S-matrix form}.

In contrast to the cylinder, the reflection amplitude on the strip, $R_b(p)$  vanishes for a discrete set of purely imaginary momenta, 
$p=i\th$, for $\theta<0$, which correspond to bound states.  The analytical continuation 
of the asymptotic fields in this region is given by complex `free fields', whose real and imaginary parts are related to each other. 
It was argued in \cite{Dorn:2008sw} that the analytical continuation of the $S$-matrix to the bound states defines 
the scalar product between the Fock space vectors of this sector.

To compute the chiral part of the $S$-matrix we have commented on the canonical structure and used it to determine the 
action $S(\varphi)$ of a one-dimensional theory, which has a non-local first-order kinetic energy term and an exponential potential. 
It reproduces the scattering amplitudes in the chiral sector, both semiclassical and quantum, via a loop expansion in 
Feynman diagrams, where the theory is regularised simply by 'normal ordering' the potential. 
More explicitly, we propose the equation 
\be
\ba
%\sum_{\nu\geq 2}{1\over\nu!}\int_{0}^{2\pi}\prod_{i=1}^\nu{dx_i\over 2\pi}
%\int D\varphi\, e^{-{1\over\hbar}S(\varphi)}\varphi(x_i)\cdots\varphi(x_\nu)\,j(x_1)\cdots j(x_\nu)
\int\!\! D\varphi\,e^{-\f{1}{\hbar}(S(\varphi)-\langle\varphi j\rangle)}
=e^{{i\over\hbar}\tilde F_q(j)}
\ea
\ee
with $\tilde F_q$ defined in \eqref{Fqtilde} and the 
sources $j$ are related to the modes of the Liouville scattering problem 
by a simple rescaling as in \eqref{aJ}.  
We proved this at the semi-classical, i.e. tree  
level and performed some non-trivial checks at the loop level.  
We consider the identification of the action $S$ as the main result of this paper. It would be interesting 
to study it further with the aim of computing the complete 
quantum generating functional.

\vspace{9mm}

\noindent
\textbf{\large Acknowledgements}
\medskip

\noindent
We acknowledge helpful discussions with Enrico Brehm,  Anamaria Font, Antal Jevicki,  Axel Kleinschmidt and Evgeny Skvortsov 
and thank Harald Dorn for discussions and comments on the manuscript.  
A part of G. J.'s work was done during his visit at Brown University, which was supported by a Fulbright Fellowship. 
His work was also supported by the joint grant of Volkswagen Foundation and SRNSF (Ref. 93 562 \& \#04/48).

\newpage

\appendix

\setcounter{equation}{0}
\def\theequation{A.\arabic{equation}}

\section{Chiral symplectic forms}\label{AppA}

The Liouville field and its canonically conjugated momentum are
\bea\label{A, bar A parameterisation}
&&\Phi(\tau,\s)=\f{1}{2}\,\log A'(x)+\f{1}{2}\,\log\bar A'(\bar x)-\log\left[1+\m^2\,A(x)\,\bar A(\bar x)\right]~,\\[1mm] \label{Pi=}
&&\Pi(\tau,\s)=\f 1 2\,\f{A''(x)}{A'(x)}+\f 1 2\,\f{\bar A''(\bar x)}{\bar A'(\bar x)}-\m^2\,\f{A'(x)\,\bar A(\bar x)
+A(x)\,\bar A'(\bar x)}{1+\m^2\,A(x)\,\bar A(\bar x)}~,
\eea
where $A$ and $\bar{A}$ are the screening charges of the $in$-field
(see \eqref{V-general}-\eqref{Screening charge}).

The canonical 2-form $\Omega$ defined by the left hand side of \eqref{Can 2-form} can be written as
\be\label{Can 2-form=}
\Omega=\Omega_0+\Omega_1+\Omega_2+\Omega_3~,
\ee
with
\be\label{Omega0}
\Omega_0=\int_0^{2\pi}\f{\text{d}\s}{2\pi}\,\left[\left(\f{{\d} A'}{2A'}\right)' \wedge\f{{\d} A'}{2A'}
+\left(\f{{\d}\bar A'}{2\bar A'}\right)' \wedge\f{{\d}\bar A'}{2\bar A'}\right]+
\f{1}{8\pi}\,\f{{\d} A'}{A'}\wedge\f{{\d}\bar A'}{\bar A'}\,
\Big |_{\s=0}^{\s=2\pi},~~~~
\ee
\be\ba\label{Omega1}
\Omega_1=\int_0^{2\pi}\f{\text{d}\s}{2\pi}\left[\f{\m^2}{2}\left(\f{{\d} A'}{A'}-\f{{\d}\bar A'}{\bar A'}\right)\wedge 
{\d}\left(\f{A'\bar A-A\bar A'}{1+\m^2\, A\,\bar A}\right)\right]
~~~~~~~~~~~~~~~~~~~~~~~~~~~\\
-\f{1}{4\pi}\,\left(\f{{\d} A'}{A'}-\f{{\d}\bar A'}{\bar A'}\right)\wedge{\d}\left(\log[1+\m^2\, A\,\bar A]\right)\,\Big |_{\s=0}^{\s=2\pi},
\ea\ee
\be\label{Omega2}
\Omega_2=\int_0^{2\pi}\f{\text{d}\s}{2\pi}\left[\f{\m^2}{2}\left(\f{{\d} A'}{A'}+\f{{\d}\bar A'}{\bar A'}\right)\wedge 
{\d}\left(\f{A'\bar A+A\bar A'}{1+\m^2\, A\,\bar A}\right)\right]~,
~~~~~~~~~~~~~~~~~~~~~~~~~~~
\ee
\be\ba\label{Omega3}
\Omega_3=\int_0^{2\pi}\f{\text{d}\s}{2\pi}\left[
\f{\m^4\left(\bar A\,{\d} A'+A\,{\d}\bar A'\right)\wedge{\d}\left(A\,\bar A\right)}{\left(1+\m^2\,A\,\bar A\right)^2}
+\m^2\,\,\f{{\d}\bar A\wedge{\d} A'-{\d}\bar A'\wedge{\d} A}{1+\m^2\,A\,\bar A}\right]\\
+\f{\m^2}{2\pi}\,\f{{\d} A\wedge{\d}\bar A}{1+\m^2\, A\,\bar A}\,\Big |_{\s=0}^{\s=2\pi}.
\ea\ee
Here, we use the relations $A'(x)=\p_\s A(x)$, $ \bar A'(\bar x)=-\p_\s \bar A(\bar x)$ and apply partial integrations 
in $\Omega_1$ and $\Omega_3$. This extracts the boundary terms in \eqref{Omega1} and \eqref{Omega3}.

The integrands of $\Omega_1$, $\Omega_2$ and $\Omega_3$
cancel each other in \eqref{Can 2-form=}. Using the monodromy of the screening charges
$A(x+2\pi)=e^{2\pi p}\,A(x)$, $\bar A(\bar x+2\pi)=e^{2\pi p}\,\bar A(\bar x)$, one finds that the boundary 
terms of $\Omega_1$ and $\Omega_3$ cancel each other as well.

Thus, $\Omega=\Omega_0=\o+\bar\o$, with similar chiral and antichiral parts. After the shift of the integration
variable $\s=x-\tau$, one finds 
\be\label{omega-A}
\o=\int_\tau^{\tau+2\pi}\f{\text{d}x}{2\pi}\,\left[
\left(\f{{\d} A'(x)}{2A'(x)}\right)' \wedge\f{{\d} A'(x)}{2A'(x)}\right]
+\f{1}{2}\,{\d} p\wedge\f{{\d} A'(\tau)}{2A'(\tau)}~.
\ee
This 2-form is $\tau$-independent and in terms of the $in$-field \eqref{chiral parts} one obtains
\eqref{omega-in}.

\setcounter{equation}{0}
\def\theequation{B.\arabic{equation}}

\section{Generating function as action functional}\label{AppB}

In Section \ref{Canonical Structure} we have introduced the generating functionals $G$ and $\bar G$. 
We now describe a different realisation of the generating functions based on the calculation of the action functional on 
the solutions of the dynamical equations. 

Let us consider a Liouville field $\Phi(\tau,\s)$ and introduce two free fields $\Phi_{-}(\tau,\sigma)$ and $\Phi_{+}(\tau,\s)$, 
which are tangent to $\Phi(\tau,\sigma)$ at $\tau=\tau_-$ and $\tau=\tau_+$, respectively. We also introduce the actions 
$S[\tau_+,\tau_-]$, $\,\,S_-[\tau_-,\tau_0]$
and $\,S_+[\tau_0,\tau_+]$,   calculated for $\Phi$, $\Phi_{-}$ and $\Phi_{+}$ respectively, for the corresponding time intervals, i.e
\be\ba\label{S[+,-]}
&S[\tau_+,\tau_-]=\frac{1}{2}\int_{\tau_-}^{\tau_+}\text{d}\tau\int_0^{2\pi}\frac{\text{d}\s}{2\pi}
\left[(\p_\tau\Phi)^2-(\p_\s\Phi)^2-4\m^2\,e^{2\Phi}\right],\\[1mm]
&S_-[\tau_-,\tau_0]=\frac{1}{2}\int_{\tau_0}^{\tau_-}\text{d}\tau\int_0^{2\pi}\frac{\text{d}\s}{2\pi}\left[(\p_\tau\Phi_-)^2
-(\p_\s\Phi_-)^2\right],\\[1mm]
&S_+[\tau_0, \tau_+]=\frac{1}{2}\int_{\tau_+}^{\tau_0}\text{d}\tau\int_0^{2\pi}\frac{\text{d}\s}{2\pi}\left[(\p_\tau\Phi_+)^2-(\p_\s\Phi_+)^2\right].
\ea\ee
Using the equations of motion for $\Phi$ and $\Phi_\pm$, together with the identity
\be\label{quadratic}
(\p_\tau\Phi)^2-(\p_\s\Phi)^2=\p_\tau(\Phi\p_\tau\Phi)-\p_\s(\Phi\p_\s\Phi)-\Phi(\p^2_\tau-\p^2_\s)\Phi~,
\ee
one obtains 
\be\ba\label{S[+,-]=}
&S[\tau_+,\tau_-]=I+\frac{1}{2}\int_0^{2\pi}\frac{\text{d}\s}{2\pi}\left[
\Phi(\tau_+,\s)\p_\tau\Phi(\tau_+,\s)- \Phi(\tau_-,\s)\p_\tau\Phi(\tau_-,\s) \right],\\[1mm]
&S_-[\tau_-,\tau_0]=\frac{1}{2}\int_0^{2\pi}\frac{\text{d}\s}{2\pi}\left[
\Phi_-(\tau_-,\s)\p_\tau\Phi_-(\tau_-,\s)- \Phi_-(\tau_0,\s)\p_\tau\Phi(\tau_0,\s)
\right],\\[1mm]
&S_+[\tau_0, \tau_+]=\frac{1}{2}\int_0^{2\pi}\frac{\text{d}\s}{2\pi}\left[
\Phi_+(\tau_0,\s)\p_\tau\Phi_+(\tau_0,\s)- \Phi_+(\tau_+,\s)\p_\tau\Phi(\tau_+,\s)
\right],
\ea\ee
with 
\be\label{I}
I(\tau_+,\tau_-)=\int_{\tau_-}^{\tau_+}\text{d}\tau\int_0^{2\pi}\frac{\text{d}\s}{2\pi}
\left[-\f{1}{2}\,\Phi\left(\p^2_\tau-\p^2_\s\right)\Phi-2\m^2\,e^{2\Phi}\right].
\ee
Since the fields $\Phi_{\pm}$ are tangent to the Liouville field $\Phi$, 
from \eqref{S[+,-]=} we find
\be\ba\label{G integral}
\lim_{t_{\pm\rightarrow\pm\infty}}&\left(S_0[\tau_0,\tau_+]+S[\tau_+,\tau_-]+S_0[\tau_-,\tau_0]\right)\\[1mm]
&=2\m^2 \int_{-\infty}^{\infty}\text{d}\tau\int_0^{2\pi}\frac{\text{d}\s}{2\pi}\,e^{2\Phi}\left(\Phi-1\right)\\[1mm]
&\qquad\qquad+\frac{1}{2}\int_0^{2\pi}\frac{\text{d}\s}{2\pi}\left[
\Phi_{\text{out}}(\tau_0,\s)\p_\tau\Phi_{\text{out}}(\tau_0,\s)- \Phi_{\text{in}}(\tau_0,\s)\p_\tau\Phi_0(\tau_0,\s)\right].
\ea\ee
The left hand side here corresponds to the generating function \eqref{cal G=} at $\tau=\tau_0$. This leads to 
\be\label{I=G+bar G}
2\m^2 \int_{-\infty}^{\infty}\text{d}\tau\int_0^{2\pi}\frac{\text{d}\s}{2\pi}\,e^{2\Phi}\left(\Phi-1\right)=G+\bar G~,
\ee
where $G$ is the generating function \eqref{G=} and $\bar G$ is its antichiral counterpart.


\newpage

\begin{thebibliography}{Ref}

%\cite{Polyakov:1981rd}
\bibitem{Polyakov:1981rd}
  A.~M.~Polyakov,
  ``Quantum Geometry of Bosonic Strings,''
  Phys.\ Lett.\ B {\bf 103} (1981) 207
  %doi:10.1016/0370-2693(81)90743-7
%CITATION = doi:10.1016/0370-2693(81)90743-7;%%

%\cite{Teschner:2001rv}
\bibitem{Teschner:2001rv}
  J.~Teschner,
  ``Liouville theory revisited,''
  Class.\ Quant.\ Grav.\  {\bf 18} (2001) R153
  %doi:10.1088/0264-9381/18/23/201
  [hep-th/0104158]
%CITATION = doi:10.1088/0264-9381/18/23/201;%%
  
%\cite{Nakayama:2004vk}
\bibitem{Nakayama:2004vk}
  Y.~Nakayama,
  ``Liouville field theory: A Decade after the revolution,''
  Int.\ J.\ Mod.\ Phys.\ A {\bf 19} (2004) 2771
%doi:10.1142/S0217751X04019500
[hep-th/0402009]
%%CITATION = doi:10.1142/S0217751X04019500;%%
  
%\cite{Seiberg:1990eb}
\bibitem{Seiberg:1990eb}
  N.~Seiberg,
  ``Notes on quantum Liouville theory and quantum gravity,''
  Prog.\ Theor.\ Phys.\ Suppl.\  {\bf 102} (1990) 319
%doi:10.1143/PTPS.102.319%\cite{Nambu:1968rr}
  
%\cite{Levy:2018bdc}
\bibitem{Levy:2018bdc}
  T.~Levy and Y.~Oz,
  ``Liouville Conformal Field Theories in Higher Dimensions,''
  JHEP {\bf 1806} (2018) 119
  %doi:10.1007/JHEP06(2018)119
  [arXiv:1804.02283 [hep-th]]
%%CITATION = doi:10.1007/JHEP06(2018)119;%%

%\cite{Zamolodchikov:1995aa}
\bibitem{Zamolodchikov:1995aa}
  A.~B.~Zamolodchikov and A.~B.~Zamolodchikov,
  ``Structure constants and conformal bootstrap in Liouville field theory,''
  Nucl.\ Phys.\ B {\bf 477} (1996) 577
  %doi:10.1016/0550-3213(96)00351-3
  [hep-th/9506136]
%%CITATION = doi:10.1016/0550-3213(96)00351-3;%%

%\cite{Balog:1997zz}
\bibitem{Balog:1997zz}
  J.~Balog, L.~Feher and L.~Palla,
  ``Coadjoint orbits of the Virasoro algebra and the global Liouville equation,''
  Int.\ J.\ Mod.\ Phys.\ A {\bf 13} (1998) 315
 %doi:10.1142/S0217751X98000147
 [hep-th/9703045]
  %%CITATION = doi:10.1142/S0217751X98000147;%%
  
\bibitem{Liouville}
J.~Liouville,
J. Math. Pures Appl., 18, 71 (1853)

%\cite{Nambu:1968rr}
\bibitem{Nambu}
Y.~Nambu,
``$S$-Matrix in semiclassical approximation,''
Phys. Lett. B \textbf{26} (1968) 626
%doi:10.1016/0370-2693(68)90436-X
  
%\cite{Boulware:1968zz}
\bibitem{BB}
D.~G.~Boulware and L.~S.~Brown,
``Tree Graphs and Classical Fields,''
Phys. Rev. \textbf{172} (1968) 1628
%doi:10.1103/PhysRev.172.1628

%\cite{Dorn:1994xn}
\bibitem{Dorn:1994xn}
H.~Dorn and H.~J.~Otto,
``Two and three point functions in Liouville theory,''
Nucl.\ Phys.\ B {\bf 429} (1994) 375
%doi:10.1016/0550-3213(94)00352-1
[hep-th/9403141],\\
%%CITATION = doi:10.1016/0550-3213(94)00352-1;%%
%\cite{Dorn:1992xw}
%\bibitem{Dorn:1992xw}
H.~Dorn and H.~J.~Otto,
``Remarks on the continuum formulation of noncritical strings,''
hep-th/9212004
%%CITATION = HEP-TH/9212004;%%
%3 citations counted in INSPIRE as of 02 Jan 2019
  
%\cite{Jorjadze:2001nx}
\bibitem{Jorjadze:2001nx}
G.~Jorjadze and G.~Weigt,
``Poisson structure and Moyal quantization of the Liouville theory,''
Nucl.\ Phys.\ B {\bf 619} (2001) 232
% doi:10.1016/S0550-3213(01)00525-9
[hep-th/0105306]
%%CITATION = doi:10.1016/S0550-3213(01)00525-9;%%

%\cite{Dorn:2008sw}
%\bibitem{Dorn:2008sw}
%H.~Dorn and G.~Jorjadze,
%``Operator Approach to Boundary Liouville Theory,''
%Annals Phys.\  {\bf 323} (2008) 2799
% doi:10.1016/j.aop.2008.02.009
%[arXiv:0801.3206 [hep-th]],\\
%%CITATION = doi:10.1016/j.aop.2008.02.009;%%
%\cite{Dorn:2006ys}
%\bibitem{Dorn:2006ys}
%H.~Dorn and G.~Jorjadze,
%``Boundary Liouville theory: Hamiltonian description and quantization,''
%SIGMA {\bf 3} (2007) 012
%doi:10.3842/SIGMA.2007.012
%[hep-th/0610197]
%%CITATION = doi:10.3842/SIGMA.2007.012;%%

%\cite{Braaten:1982yn}
\bibitem{Braaten:1982yn}
%\cite{Curtright:1982gt}
%\bibitem{Curtright:1982gt}
T.~L.~Curtright and C.~B.~Thorn,
``Conformally Invariant Quantization of the Liouville Theory,''
Phys. Rev. Lett. \textbf{48} (1982) 1309
[erratum: Phys. Rev. Lett. \textbf{48} (1982) 1768];\\
%doi:10.1103/PhysRevLett.48.1309
%342 citations counted in INSPIRE as o  
E.~Braaten, T.~Curtright and C.~B.~Thorn,
``An Exact Operator Solution of the Quantum Liouville Field Theory,''
Annals Phys.\  {\bf 147} (1983) 365
% doi:10.1016/0003-4916(83)90214-2
%%CITATION = doi:10.1016/0003-4916(83)90214-2;%%
%181 citations counted in INSPIRE as of 02 Jan 2019  
  
%\cite{Otto:1985eb}
\bibitem{Otto:1985eb}
H.~J.~Otto and G.~Weigt,
``Construction Of Exponential Liouville Field Operators For Closed String Models,''
Z.\ Phys.\ C {\bf 31} (1986) 219
%doi:10.1007/BF01479530
%%CITATION = doi:10.1007/BF01479530;%%
%54 citations counted in INSPIRE as of 02 Jan 2019  

%\cite{Gervais:1983ry}
\bibitem{Gervais:1983ry}
J.~L.~Gervais and A.~Neveu,
``Novel Triangle Relation and Absence of Tachyons in Liouville String Field Theory,''
Nucl.\ Phys.\ B {\bf 238} (1984) 125
%doi:10.1016/0550-3213(84)90469-3
%%CITATION = doi:10.1016/0550-3213(84)90469-3;%%
%288 citations counted in INSPIRE as of 02 Jan 2019  

%\cite{Fateev:2000ik}
\bibitem{Fateev:2000ik}
V.~Fateev, A.~B.~Zamolodchikov and A.~B.~Zamolodchikov,
``Boundary Liouville field theory. 1. Boundary state and boundary two point function,''
hep-th/0001012
%%CITATION = HEP-TH/0001012;%%
%310 citations counted in INSPIRE as of 02 Jan 2019 
  
%\cite{Ponsot:2001ng}
\bibitem{Ponsot:2001ng}
B.~Ponsot and J.~Teschner,
``Boundary Liouville field theory: Boundary three point function,''
Nucl.\ Phys.\ B {\bf 622} (2002) 309
%doi:10.1016/S0550-3213(01)00596-X
[hep-th/0110244]
%%CITATION = doi:10.1016/S0550-3213(01)00596-X;%%
%108 citations counted in INSPIRE as of 02 Jan 2019  
  
%\cite{Dorn:2008sw}
\bibitem{Dorn:2008sw}
H.~Dorn and G.~Jorjadze,
``Operator Approach to Boundary Liouville Theory,''
Annals Phys.\  {\bf 323} (2008) 2799
% doi:10.1016/j.aop.2008.02.009
[arXiv:0801.3206 [hep-th]]
%%CITATION = doi:10.1016/j.aop.2008.02.009;%%
%5 citations counted in INSPIRE as of 02 Jan 2019


\end{thebibliography}
\end{document}